\documentclass[twocolumn,superscriptaddress,floatfix,prb,amsmath,aps,showpacs]{revtex4}
\usepackage{graphicx}
\usepackage{bm}
\usepackage{amssymb}
\usepackage[dvips]{color}
\begin{document}
\bibliographystyle{apsrev}
\def\nn{\nonumber}
\def\dag{\dagger}
\def\u{\uparrow}
\def\d{\downarrow}
\def\j{\bm j}
\def\m{\bm m}
\def\l{\bm l}
\def\0{\bm 0}
\def\k{\bm k}
\title{chiral spin-wave edge modes in dipolar magnetic thin films}
\date{\today}
\author{Ryuichi Shindou} 
\affiliation{Department of Physics, Tokyo Institute of Technology,
2-12-1 Ookayama, Meguro-ku, Tokyo, 152-8551, Japan} 
\affiliation{International Center for Quantum Materials, Peking 
University, No.5 Yiheyuan Road, Haidian District, Beijing, 100871, China} 
\author{Jun-ichiro Ohe}
\affiliation{Department of Physics, Toho University, 
2-2-1 Miyama, Funabashi, Chiba, Japan}
\author{Ryo Matsumoto} 
\affiliation{Department of Physics, Tokyo Institute of Technology,
2-12-1 Ookayama,  Meguro-ku, Tokyo, 152-8551, Japan}  
\author{Shuichi Murakami} 
\affiliation{Department of Physics, Tokyo Institute of Technology,
2-12-1 Ookayama, Meguro-ku, Tokyo, 152-8551, Japan}  
\author{Eiji Saitoh}
\affiliation{Institute for Materials Research, Tohoku University, Sendai, 980-8557, Japan}
\begin{abstract}
Based on a linearized Landau-Lifshitz equation, 
we show that two-dimensional periodic allay of ferromagnetic 
particles coupled with magnetic dipole-dipole interactions 
supports chiral spin-wave edge modes, when subjected 
under the magnetic field applied perpendicular to the plane.
The mode propagates along a 
one-dimensional boundary of the system in a unidirectional way 
and it always has a chiral dispersion within a band gap for  
spin-wave volume modes. Contrary to the well-known 
Damon-Eshbach surface mode, 
the sense of the rotation depends not only on 
the direction of the field but also on the strength of the  
field; its chiral direction is generally determined by the sum 
of the so-called Chern integers defined for spin-wave volume 
modes below the band gap. Using simple tight-binding descriptions, 
we explain how the magnetic dipolar interaction endows 
spin-wave volume modes with non-zero Chern integers 
and how their values will be changed by the field.     
\end{abstract}
\maketitle
\section{introduction}
Spin waves are collective propagations of precessional motions 
of magnetic moments in magnetic materials. Magnonics research 
investigates how the spin wave propagates in the sub-micrometer 
length scale and sub-nanosecond  time 
scale.~\cite{Damon1,Damon2,Kalinikos,Kryuglyak,Serga,Lenk}
Especially, the propagation of spin waves in periodically nanostructured  
magnetic materials dubbed as magnonic 
crystals~\cite{Gulyaev,Singh,ccWang,Adeyeye}  
are of one of its central interests. Owing to the periodic 
structurings, the spin wave spectrum in magnonic 
crystal acquires allowed frequency bands of spin 
wave modes and forbidden-frequency bands 
dubbed as magnonic band gap.~\cite{Gulyaev} 
Like in other solid-state engineering such 
as electronics, photonics and plasmonics, the 
main application direction 
is to explore ability of spin waves to carry and 
process information. Compared to others, magnonics 
has a much better prospect for miniaturization 
of the device, because the velocity of a spin
wave is typically several orders slower than 
those of light and electrons in solids.     

Recently, the authors proposed a spin-wave analog  
of integer quantum Hall (IQH) state,~\cite{smmo} which has  
unidirectional edge modes for spin-wave propagation.  
IQH state is a two-dimensional electron system 
with broken time-reversal symmetry, which supports 
unidirectional electric conducting channels along 
the boundaries (edges) of the system.~\cite{QHE}   
The number of the unidirectional  (chiral) edge modes is 
determined by a certain kind of topological number defined  
for {\it bulk} electronic states, called as the first 
Chern integer.~\cite{TKNN,Halperin,Hatsugai}  
Based on a linearized Landau-Lifshitz equation, 
we have generalized the Chern integer well-established 
in quantum Hall physics into the context of the spin 
wave physics, to argue that non-zero Chern integer for 
spin-wave {\it volume-mode} bands results in an  
emergence of chiral spin-wave edge mode.~\cite{smmo} 

The proposed edge mode has a chiral dispersion with 
a band gap for volume-mode bands, which 
supports a unidirectional propagation of spin degree of 
freedom for a frequency within the gap. 
The sense of rotation and the number 
of the chiral mode is determined by the 
topological number for volume-mode bands 
below the gap, which itself can be changed by closing the 
band gap. These features allow us control 
the chiral edge modes in terms of band-gap 
manipulation, which could realize novel 
spintronic devices such as spin current 
splitter and spin-wave logic gates.~\cite{smmo} 
To have these devices in real experimental 
systems, however, it is quite important to have 
a number of actual magnonic crystals, in which 
spin-wave volume mode bands take various 
non-zero Chern integers. 
 
From its electronic analogue,~\cite{KL,onoda-nagaosa1} 
it is expected that non-zerzo Chern integers for  
spin-wave volume-mode bands result from strong 
spin-orbit coupled interactions, such as magnetic 
dipole-dipole interaction. Namely, having an inner product 
between spin operator and coordinate operator, the 
magnetic dipolar interaction locks the relative rotational 
angle between the spin space and orbital space, just 
in the same way as the relativistic spin-orbit interaction 
does in electronic systems.~\cite{KL,onoda-nagaosa1} 
As a result of the spin-orbit 
locking, the complex-valued character in the spin space 
(i.e. one of the three Pauli matrices) is transferred  
into wavefunctions in the orbital space. Especially, 
in the presence of finite out-of-plane ferromagnetic moments 
in the spin space, the symmetry argument allows the Chern 
integer for volume-mode bands to have non-zero 
integer-value. In the recent work, employing a 
standard plane-wave theory, we have 
showed that a two-dimensional ($x$-$y$) bi-component 
magnonic crystal under an  out-of-plane field 
(along the $z$-direction) acquires spin-wave 
volume-mode bands with non-zero Chern 
integers, when magnetic dipolar interaction 
dominates over short-ranged isotropic exchange 
interaction. From the state-of-art 
nanotechnology, however, it is not easy to 
synthesize the proposed bi-component magnonic 
crystal experimentally. Moreover, the proposed 
model is not simple enough to see how magnetic 
dipolar interaction leads to non-zero Chern integers  
for spin-wave volume-mode bands.  

In the present paper, we introduce much simpler 
thin-film magnetic models, which also support 
spin-wave volume modes with non-zero Chern integers 
and chiral spin-wave edge modes, under 
the field normal to the two-dimensional plane. 
Based on the models, we show that the chiral edge modes have  
frequency-wavelength dispersions within a band gap for 
spin-wave volume modes, and their chiral 
directions are determined by a sign of the Chern 
integer for a spin-wave volume mode below the gap. 
Using a simple tight-binding model composed of 
`atomic orbitals', we further 
argue that the level inversion between the 
parity-odd  atomic orbital (such as $p$-wave orbital) 
and parity-even atomic orbital (such as $s$-wave orbital) 
leads to a band inversion, which endows 
spin-wave volume-mode bands with 
non-zero Chern integers. We expect that these 
findings would give useful prototype models for future 
designing of more realistic magnonic crystals which 
support topological chiral spin-wave edge modes.

The organization of the paper is as follows. 
In the next two sections, we introduced the 
models studied in this paper (sec.~II and Fig.~\ref{fig:model}) 
and formulate our problem and summarize a calculation 
procedure of spin-wave band dispersions and 
the topological Chern integers (sec.~III). 
In section IV, we show how chiral  spin-wave 
edge modes appear and how they change their 
directions on increasing the field. The results shows that 
the sense of the rotation of chiral edge mode is 
indeed determined by the sign of the Chern integer defined 
for the spin-wave volume mode. In section V, we 
introduce a tight-binding description of linearized 
Landau-Lifshitz equations in the context of the 
present models. In sec.VA, we first clarify spin-wave 
excitations within a unit cell  
in terms of a total angular momentum variable. 
Based on the `atomic orbitals' thus obtained, 
we construct a tight-binding model for a 
square-lattice model (sec. VB) and for a honeycomb-lattice model 
(sec. VC).  Using this tight-binding model, 
we explain how a level inversion between different 
`atomic-orbital'  levels leads to an {\it inverted} 
spin-wave band with non-zero Chern integers 
and how the signs of the Chern integers are   
changed as a function of the field. 
To see how the proposed chiral 
spin-wave edge modes could be seen in 
experiments, we simulate the 
Landau-Lifshitz-Gilbert equation for 
the square-lattice model near the saturation 
field (sec. VI). The section VII is devoted for 
summary and future open issues, in 
which we also discuss the effects of disorders 
associated with lattice periodicity and 
shape of the boundaries.   
  
\section{Model}
In this paper, we consider two-dimensional periodic 
arrays of ferromagnetic islands. 
We assume that each ferromagnetic island 
behaves as a single spin and ferromagnetic 
islands are coupled via magnetic dipolar interaction. 
In fact, two-dimensional periodic lattice structures 
composed of submicrometer-scale ferromagnetic islands 
have been fabricated experimentally, in which they confirm 
that each island behaves as a giant single spin 
under some circumstances.~\cite{asi,CARoss} 

\begin{figure}
   \includegraphics[width=60mm]{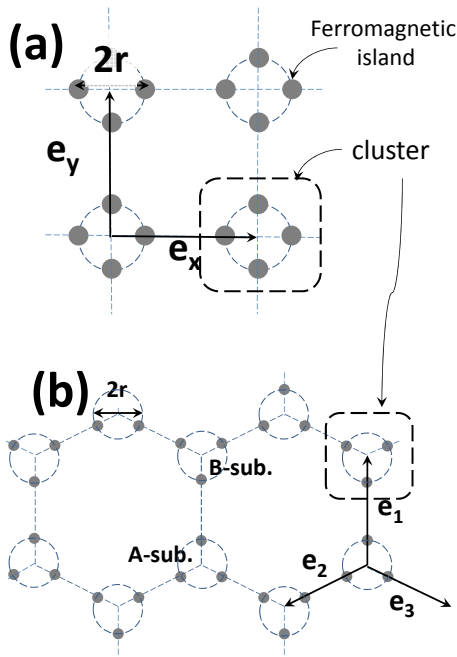}
\caption{ (Color online) Periodic array of 
ferromagnetic islands (gray circles) (a) 
decorated square-lattice model. (b) decorated 
honeycomb-lattice model. Each gray point 
stands for a ferromagnetic island (volume element is 
$\Delta V$), which 
we assume to behaves as a single big spin whose 
moment is fully saturated ($M_s$). 
The spins are coupled via magnetic 
dipole-dipole interaction. We took 
$|{\bm e}_x|=|{\bm e}_y|=2.4$, $2r=1.2$, 
$\Delta V=1.70$ and $M_s=1.0$
for the square-lattice case, while 
$|{\bm e}_1|=|{\bm e}_2|=|{\bm e}_3|=2.4$, $2r=1.2$, 
$\Delta V=1.0$ and $M_s=1.0$ for 
the honeycomb-lattice case. The primitive translational 
vector ${\bm a}_{\mu}$ ($\mu=x,y$) are defined  
as ${\bm a}_x={\bm e}_x$ and ${\bm a}_y={\bm e}_y$ 
for the square-lattice model and as 
${\bm a}_x={\bm e}_3-{\bm e}_2$ and 
${\bm a}_y={\bm e}_1-{\bm e}_2$ for the 
honeycomb-lattice case (see text). }
\label{fig:model}
\end{figure}

To have volume-mode bands with finite Chern integers, 
we generally need multiple-band degree of freedom within 
a unit cell of magnonic crsytal. To this end, we 
consider two models; 
decorated square-lattice model and honeycomb lattice model 
(see Fig.~\ref{fig:model}). A basic building block of both 
models is a cluster of ferromagnetic islands. 
For the decorated square lattice 
model, four ferromagnetic islands form a circle-shape 
cluster which encompasses a site of the square lattice.  
For the decorated honeycomb lattice model, $3$ neighboring 
islands form a circle which encompasses 
either an A-sublattice site 
or a B-sublattice site of the honeycomb lattice. 

Experimentally speaking, it is also quite likely that a 
submicrometer-scale ferromagnetic island has a 
number of low-energy excitation modes having different spin 
textures within the island. Such modes can be also regarded 
as multiple-band degree of freedom, so that a system with 
only one ferromagnetic island within a unit cell of magnonic 
crystal~\cite{CARoss} could also 
have a chance to provide volume mode bands with 
finite Chern integer and associated chiral edge modes. We 
expect that the theoretical results obtained in the present 
model study would also provide useful starting points 
for further studies on such systems.~\cite{later}

\section{Formulation}
For the models introduced above, we first determine 
a classical spin configuration which 
minimizes the following magnetostatic energy;
\begin{align}
E&= - \frac12 \!\ (\Delta V)^2 \!\ \sum^{i\ne j}_{i,j} 
M_{a}({\bm r}_i) f_{ab}({\bm r}_i-{\bm r}_j) M_{b}({\bm r}_j) \nn \\
& \ \  \ + H \Delta V \!\ \sum_{i} M_{z}({\bm r}_j). \label{magsta}
\end{align}  
where ${\bm r}_i$ specifies a spatial location of  a 
ferromagnetic island (classical spin). 
For simplicity, the norm of each spin is fixed; 
$|{\bm M}({\bm r}_j)|=M_s$. 
The magnetic dipole-dipole interaction is given by a 
3 by 3 matrix, 
\begin{align}
f_{ab}({\bm r}) = - \frac{1}{4\pi} 
\bigg(\frac{\delta_{a,b}}{|{\bm r}|^3} 
- \frac{3r_ar_b}{|{\bm r}|^5} \bigg).  
\end{align}
with $a,b=x,y,z$. The summation 
over $i,j$ in eq.~(\ref{magsta}) are taken over all 
ferromagnetic islands, while the summation 
over $a,b=x,y,z$ were omitted. 
A corresponding Landau-Lifshitz equation reads
\begin{align}
\partial_t M_a({\bm r}_i) 
&= \epsilon_{abc} \big(H \delta_{b,z} \nn \\
& \ \ \ \ - \Delta V \!\ \sum_{j \ne i} f_{bd}({\bm r}_i-{\bm r}_j) M_{d}({\bm r}_j) 
\big) M_{c}({\bm r}_i). \label{LL} 
\end{align}
$\Delta V$ is a volume element for each ferromagnetic 
island. From dimensional analysis, one can 
see that a saturation field and resonance frequency 
of spin-wave excitations are scaled by $M_s \Delta V/l^3$, 
where $l$ is a characteristic length scale for 
the periodic structuring within the two-dimensional 
plane, e.g. radius ($r$) of the circle-shape $M$-spin cluster. 
In the following, we take this value to be 
around $1$; $M_s=1$, $\Delta V=1.0$,  
$2r=1.2$ for the square-lattice case and 
$M_s=1$, $\Delta V=1.7$, $2r=1.2$ for the 
honeycomb-lattice case. 

\subsection{classical spin configuration}
\subsubsection{square-lattice model}
\begin{figure}
   \includegraphics[width=80mm]{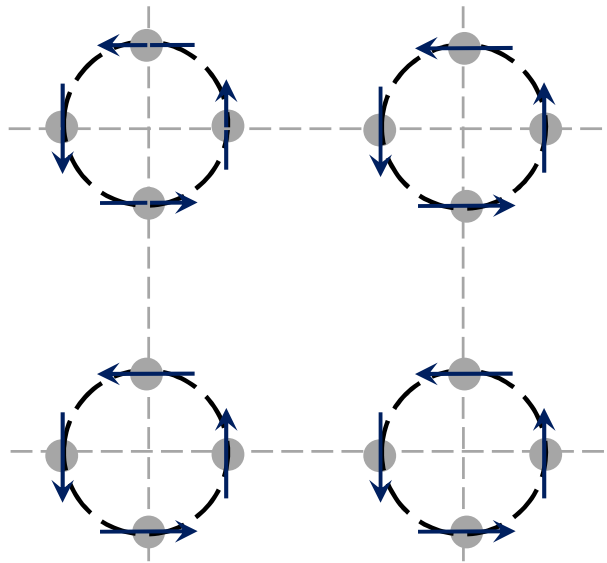}
\caption{ (Color online) Top-view of classical spin 
configurations for the decorated square-lattice model.
The field is lower than the saturation field, so that  
spins have finite in-plane components, forming a 
vortex structure.}
\label{fig:classical}
\end{figure}

For the decorated square lattice case, we found that 
every four spins within a circle-shape cluster  
form a same vortex, 
\begin{eqnarray}
M_0({\bm r}=r(c_{\theta_j},s_{\theta_j})) 
= M_s (-s_{\varphi}s_{\theta_j},s_{\varphi}c_{\theta_j},c_{\varphi}), \label{csc}
\end{eqnarray}
with $\theta_j\equiv \frac{2\pi j}{4}$ ($j=1,\cdots,4$) 
and $(s_{\theta},c_{\theta})\equiv (\sin\theta,\cos\theta)$, 
such that the classical spin configuration ${\bm M}_0({\bm r})$ 
respects the translational symmetries 
of the square lattice, 
${\bm M}_{0}({\bm r}+{\bm a}_{\mu})={\bm M}_{0}({\bm r})$ 
(see Fig.~\ref{fig:classical}). 
A finite out-of-plane component ($\varphi\neq \frac{\pi}{2}$) 
is induced by the field. Above the saturation field 
($H>H_{c}= 1.71$), all the spins become fully polarized 
along the field ($\varphi=\pi$). 
\subsubsection{honeycomb-lattice model}
For the decorated honeycomb 
lattice case, the classical spin configuration below a saturation field 
($H<H_{c}=0.57$) breaks the 
translational symmetries of the lattice, while that above the 
field is a fully polarized state respecting the translational  
symmetries of the honeycomb lattice. For simplicity, we only 
consider spin-wave excitations above the saturation field 
for the decorated honeycomb lattice case.

\subsection{linearized Landau-Lifshitz equation}
Starting from the classical spin configurations thus obtained, 
${\bm M}_0({\bm r})$, the Landau-Lifshitz equation 
is linearized with respect to a small transverse 
fluctuation field ${\bm m}_{\perp}({\bm r})$, with
${\bm M}({\bm r}) \equiv {\bm M}_{0}({\bm r}) 
+ {\bm m}_{\perp}({\bm r})$ and 
${\bm m}_{\perp}\perp {\bm M}_0$. In terms of a rotated frame 
with a $3$ by $3$ rotational matrix ${\bm R}({\bm r})$, 
with which ${\bm M}_{0}({\bm r})$ is always pointing along the
$z$-direction, 
${\bm R}({\bm r}) {\bm M}_0({\bm r}) \equiv M_s {\bm e}_z$ 
and ${\bm R}({\bm r}) {\bm m}_{\perp}({\bm r}) 
\equiv {\bm m}({\bm r})$,  
the linearized equation of motion for the 
transverse moments takes the form;
\begin{align}
- \partial_t m_{\mu}({\bm r}_i) &= 
\epsilon_{\mu\nu} \alpha({\bm r}_i) 
m_{\nu}({\bm r}_i) \nn \\
&\hspace{0.2cm} + M_s \Delta V \!\ \epsilon_{\mu\nu} \sum_{j\ne i} 
f_{\nu\lambda}({\bm r}_i,{\bm r}_j) m_{\lambda}({\bm r}_j) \label{swh}
\end{align}
where ${\bm m}\equiv(m_x,m_y,0)$ and the summation 
over the repeated indices $\mu,\nu,\lambda$ 
are taken only over $x,y$ with 
$\epsilon_{xy}=-\epsilon_{yx}=1$. The 
first term in the right hand side includes a  
demagnetization field and the external field;
\begin{align}
\alpha({\bm r}_i) {\bm M}_{0}({\bm r}_i) 
= -  \Delta V\!\ 
\sum_{j \ne i}{\bm f}({\bm r}_i-{\bm r}_j) 
{\bm M}_{0}({\bm r}_j) + H {\bm e}_z, \nn 
\end{align}  
where, provided that ${\bm M}_0({\bm r}_j)$ 
gives a local minimum for the magnetostatic 
energy Eq.~(\ref{magsta}), 
the equality always holds true for a certain 
scalar function $\alpha({\bm r}_i)$. 
The dipole-dipole interaction in the second term 
of eq.~(\ref{swh})
is given in the rotated frame; 
\begin{align}
{\bm f}({\bm r}_i,{\bm r}_j) \equiv 
{\bm R}({\bm r}_i) {\bm f}({\bm r}_i-{\bm r}_j) 
 {\bm R}^{t}({\bm r}_j). \nn
\end{align}  
In terms of $m_{\pm} \equiv m_{x}\pm i m_y$, which are 
magnon creation/annihilation fields respectively, 
the equation of motion 
reduces to a following form;
\begin{align}
-i\partial_t {\bm \sigma}_3 \left(\begin{array}{c}
m_{+}({\bm r}_i) \\
m_{-}({\bm r}_i) \\
\end{array}\right) &= 
\alpha({\bm r}_i) M_s \left(\begin{array}{c}
m_{+}({\bm r}_i) \\
m_{-}({\bm r}_i) \\
\end{array}\right) \nn \\ 
&\hspace{-2.8cm} + M_s \Delta V\!\ 
\sum_{j \ne i} \left(\begin{array}{cc}
f_{++}({\bm r}_i,{\bm r}_j) & f_{+-}({\bm r}_i,{\bm r}_j) \\
f_{-+}({\bm r}_i,{\bm r}_j) & f_{--}({\bm r}_i,{\bm r}_j) \\
\end{array}\right) \left(\begin{array}{c}
m_{+}({\bm r}_j) \\
m_{-}({\bm r}_j) \\
\end{array}\right), \label{eom0}
\end{align} 
where a $2$ by $2$ diagonal 
Pauli matrix ${\bm \sigma}_3$ takes 
$+1$ for the creation field (particle space), 
while take $-1$ for the annihilation field 
(hole space). A Green function $f_{\alpha\beta}({\bm r},{\bm r}')$ 
($\alpha,\beta=\pm$)  in the second term takes a form of a 
certain Hermite matrix in the  
particle-hole space;
\begin{align}
& \left(\begin{array}{cc}
f_{++}({\bm r},{\bm r}') & f_{+-}({\bm r},{\bm r}') \\
f_{-+}({\bm r},{\bm r}') & f_{--}({\bm r},{\bm r}') \\
\end{array}\right) 
= \nn \\
&\hspace{0.6cm} 
\frac12 \left(\begin{array}{cc}
1 & i \\
1 & -i \\ 
\end{array}\right) \left(\begin{array}{cc}
f_{xx}({\bm r},{\bm r}') & f_{xy}({\bm r},{\bm r}') \\
f_{yx}({\bm r},{\bm r}') & f_{yy}({\bm r},{\bm r}') \\
\end{array}\right) \left(\begin{array}{cc}
1 & 1 \\
-i & i \\ 
\end{array}\right), \nn
\end{align}
with $f^{*}_{\alpha\beta}({\bm r},{\bm r}')=
f_{\beta\alpha}({\bm r}',{\bm r})$. 
Accordingly, the problem reduces to solving 
a following generalized eigenvalue problem; 
\begin{align}
\sum_{j} ({\bm H})_{{\bm r}_i,{\bm r}_j} 
\left(\begin{array}{c}
m_{+}({\bm r}_j) \\
m_{-}({\bm r}_j) \\
\end{array}\right) = {\bm \sigma}_3 \left(\begin{array}{c}
m_{+}({\bm r}_i) \\
m_{-}({\bm r}_i) \\
\end{array}\right) \overline{E} \label{gev}
\end{align}
with an Hermite matrix ${\bm H}$, 
\begin{align}
({\bm H})_{{\bm r}_i,{\bm r}_j} 
& = -M_s \alpha({\bm r}_i) \delta_{{\bm r}_i,{\bm r}_j} 
\left(\begin{array}{cc}
1 & \\
& 1\\
\end{array}\right) \nn \\ 
&\hspace{-0.5cm} - M_s \Delta V (1-\delta_{{\bm r}_i,{\bm r}_j})\!\ 
\left(\begin{array}{cc}
f_{++}({\bm r}_i,{\bm r}_j) & f_{+-}({\bm r}_i,{\bm r}_j) \\
f_{-+}({\bm r}_i,{\bm r}_j) & f_{--}({\bm r}_i,{\bm r}_j) \\
\end{array}\right), \label{hori}
\end{align}
The sum of $j$ is taken over all spins in the systems.
Using the Cholesky decomposition,~\cite{Colpa} the 
Hermite matrix can be diagonalized by a paraunitary 
transformation matrix ${\bm T}$;
\begin{align}
{\bm H} {\bm T} = {\bm \sigma}_3 {\bm T} \overline{\bm E} \label{para}
\end{align}
with a proper normalization condition 
${\bm T}^{\dagger}{\bm \sigma}_3{\bm T}={\bm \sigma}_3$ 
and a diagonal matrix $\overline{\bm E}$.

Now that the saddle point solution respects the 
translational symmetries, ${\bm M}_0({\bm r}+{\bm a}_{\mu})
={\bm M}_0({\bm r})$, so does the Green function and 
the demagnetization coefficient, 
${\bm f}({\bm r}+{\bm a}_{\mu},{\bm r}')={\bm f}({\bm r},{\bm r}'-{\bm a}_{\mu})$ 
and $\alpha({\bm r}+{\bm a}_{\mu})=\alpha({\bm r})$   
with the primitive translational vectors ${\bm a}_{\mu}$ 
($\mu=x,y$). Moreover, the classical spin configuration 
eq.~(\ref{csc}) is invariant under the simultaneous $C_4$ rotations 
in the spin space and 
the lattice space (around $z$-axis), so that 
the demagnetization coefficient 
within a unit cell has no spatial dependence, 
$\alpha({\bm r}_j)=\alpha$. This also holds true for  
the honeycomb lattice case considered.

With the Born-von Karman 
boundary condition, the eigenvalue problem 
reduces to a diagonalization of 
following Bogoliubov-de Gennes type Hamiltonian for 
every crystal momentum ${\bm k}=(k_x,k_y)$;
\begin{align}
i\partial_t {\bm \sigma}_3 \left(\begin{array}{c}
u_{+,{\bm k}}({\bm r}_i) \\
u_{-,-{\bm k}}({\bm r}_i) \\
\end{array}\right) 
= \sum^{M_U}_{j=1}
\big({\bm H}_{\bm k}\big)_{{\bm r}_i,{\bm r}_j} 
\left(\begin{array}{c}
u_{+,{\bm k}}({\bm r}_j) \\
u_{-,-{\bm k}}({\bm r}_j) \\
\end{array}\right). \nn
\end{align}   
with 
\begin{align}
& \big({\bm H}_{\bm k}\big)_{{\bm r}_i,{\bm r}_j} 
\equiv - M_s \alpha \delta_{{\bm r}_i,{\bm r}_j}  \nn \\
&\hspace{0.6cm} 
- M_s \Delta V\!\ \left(\begin{array}{cc}
f_{{\bm k},++}({\bm r}_i,{\bm r}_j) & 
f_{{\bm k},+-}({\bm r}_i,{\bm r}_j) \\
 f_{{\bm k},-+}({\bm r}_i,{\bm r}_j) & 
f_{{\bm k},--}({\bm r}_i,{\bm r}_j) \\
\end{array}\right), \label{hori2}
\end{align} 
and 
\begin{align} 
f_{{\bm k},\alpha\beta}({\bm r},{\bm r}') &\equiv 
e^{-i{\bm k}({\bm r}-{\bm r}')} 
\sum_{\bm b} (1-\delta_{{\bm r},{\bm r}'-{\bm b}})  \nn \\
&\hspace{2cm} \times f_{\alpha\beta}({\bm r},{\bm r}'-{\bm b}) 
e^{-i{\bm k}{\bm b}}, \nn
\end{align}
and 
\begin{align}
m_{\pm}({\bm r}+{\bm a}_{\mu}) 
\equiv \sum_{\bm k} e^{\pm i {\bm k}{\bm a}_{\mu}} 
u_{\pm,{\bm k}}({\bm r}). \nn
\end{align}
The summation with respect to $j$ (or ${\bm r}_j$) in the 
right hand side is taken over a unit cell. For decorated 
square and honeycomb lattice, $M_U=4$ and $6$ respectively. 
The summation over the translation vectors 
${\bm b}$ are taken over sufficiently many unit cells 
in actual numerical calculations, 
${\bm b}=n{\bm a}_x+m{\bm a}_y$ with $-50 \le n,m \le 50$.     
In terms of the Cholesky decomposition, the 
$2M_U\times 2M_U$ BdG Hamiltonian is diagonalized 
\begin{eqnarray}
{\bm H}_{\bm k} |\psi_{j}\rangle = 
{\bm \sigma}_3 |\psi_{j} \rangle 
\overline{E}_{j,{\bm k}}. \label{para} 
\end{eqnarray} 
with the normalization condition, 
$\langle \psi_j |{\bm \sigma}_3 |\psi_j\rangle 
=(-1)^{\sigma_j}$ where $\sigma_j=0$ for particle bands,   
$j=1,\cdots,M_U$, and $\sigma_j=1$ for hole bands 
$j=M_U+1,\cdots,2M_U$. 
Provided that the spin-wave Hamiltonian is 
derived from an energy minimum of the magnetostatic 
energy Eq.~(\ref{magsta}), it is guaranteed that eigenvalues  
for particle bands ($j=1,\cdots,M$) are positive definite 
$\overline{E}_{j,{\bm k}}>0$ for any ${\bm k}$, while 
those for the hole bands ( $j=M_U+1,\cdots,2M_U$) 
are all negative, $\overline{E}_{j,{\bm k}}<0$ for 
any ${\bm k}$. In fact, this is 
true for all the cases studied in this paper.    

The eigenvalues in the particle bands, $\overline{E}_{j,{\bm k}}$ 
($j=1,\cdots,M_U$), determine wavelength-frequency dispersion 
relations for all the spin-wave volume-mode bands. An  
eigenvector, $|\psi_j\rangle$, is a    
`Bloch wavefunction' for the corresponding spin-wave 
volume-mode band. In terms 
of the Bloch wavefunction, 
we have calculated the first 
Chern integer defined for each spin-wave band as,~\cite{smmo} 
\begin{align}
{\rm Ch}_j & \equiv i (-1)^{\sigma_j} \epsilon_{\mu\nu} 
\int_{\rm BZ} d^2{\bm k} \!\ \big\langle \partial_{k_{\mu}} \psi_j 
\big| {\bm \sigma}_3 \big| \partial_{k_{\nu}} \psi_j \big\rangle \nn \\
& = i \sum^{m\ne j}_{m=1,\cdots,2M_U}  
\int_{\rm BZ} d^2{\bm k} \frac{(-1)^{\sigma_j} (-1)^{\sigma_m}}
{(\overline{E}_{j,{\bm k}}-\overline{E}_{m,{\bm k}})^2}
\nn \\ 
&\hspace{-0.5cm} 
\times \Bigg\{ \bigg\langle \psi_j 
\bigg| \frac{\partial {\bm H}_{\bm k}}{\partial k_x} \bigg| 
\psi_m \bigg\rangle
\bigg\langle \psi_m \bigg| 
\frac{\partial {\bm H}_{\bm k}}{\partial k_y} 
\bigg| \psi_{j} \bigg\rangle - (x \leftrightarrow y) \Bigg\}
 \label{ch}
\end{align}
Contrary to the Chern integer defined for electron's 
wavefunction,~\cite{TKNN} 
eq.~(\ref{ch}) contains the diagonal Pauli 
matrix ${\bm \sigma}_3$ between bra-state and 
ket-state, which takes $+1$ in the particle space 
while $-1$ in the hole space. This additional 
structure comes from the fact 
that magnon obeys the boson statistics,~\cite{smmo} 
which enforces the respective 
BdG Hamiltonian such as eqs.~(\ref{hori},\ref{hori2}) 
to be diagonalized in terms of a paraunitary matrix 
instead of a unitary matrix. Due to this paraunitary character 
in the particle-hole space, we can also argue that the 
sum of the magnonic Chern integer over all {\it particle} 
bands always reduce to zero, $\sum^{M_U}_{j=1} {\rm Ch}_j=0$, 
which leads to the absence of {\it gapless} topological 
chiral spin-wave edge mode.~\cite{smmo}

In the next section, we have calculated spin-wave 
excitations with the open boundary condition 
along one direction ($y$-direction) while the 
periodic boundary condition along the other 
($x$-dir.); the frequency-wavelength dispersions  
for the spin-wave edge modes 
are obtained as a function of (surface) 
momentum along the $x$-direction, $k_x$. The 
dispersions thus obtained alllow us to 
see the propagation direction of the chiral 
spin-wave edge mode. With changing the 
strength of the field , we have  
calculated spin-wave band dispersions for both 
volume modes and edge modes and 
the Chern integer for all the volume modes. 

\begin{figure}
\begin{center}
\includegraphics[width=96mm]{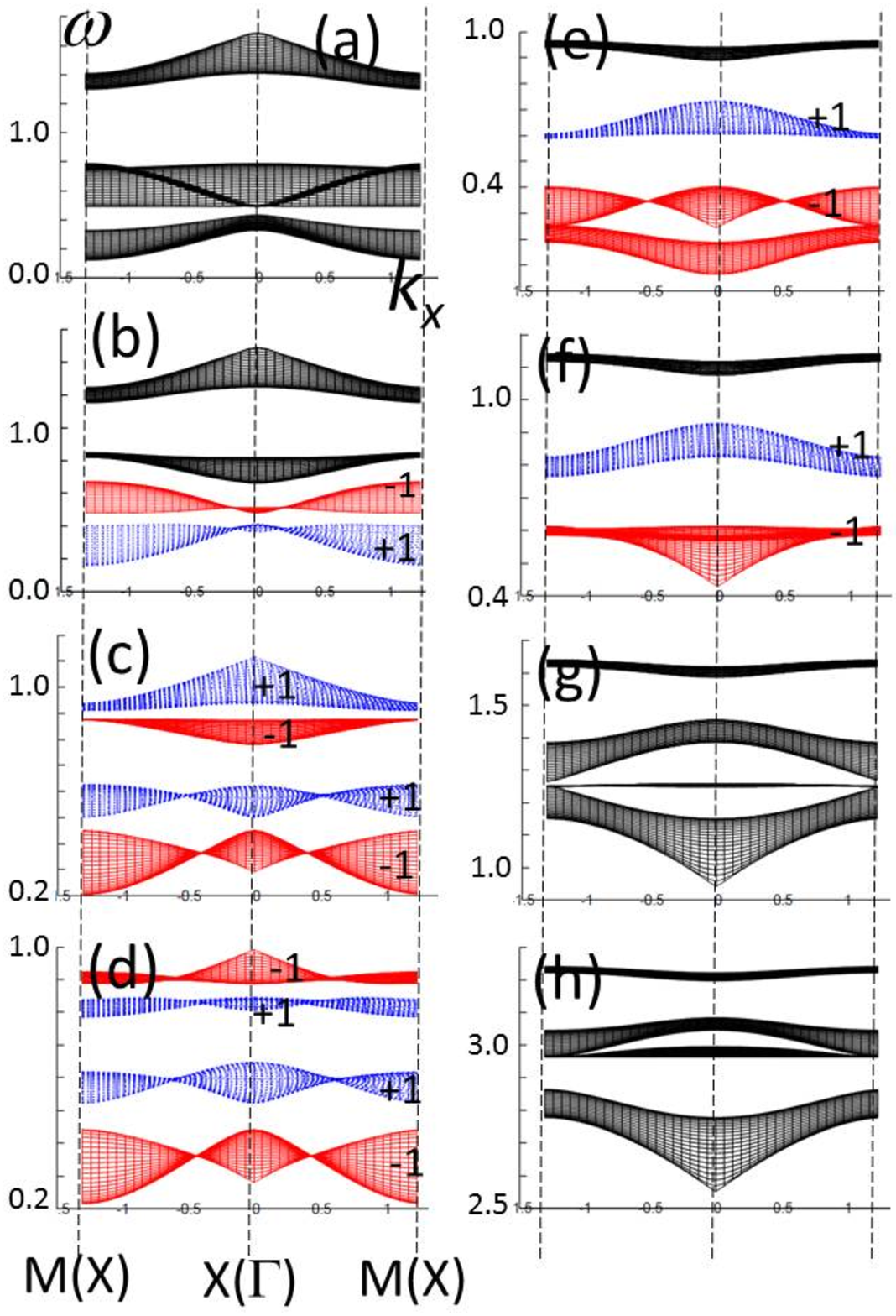}
\caption{ (Color online) (a-d) Side-view of 
wavelength-frequency dispersions of four  
spin-wave volume-mode bands in the square lattice model under 
fields ($H$) normal to the 2-$d$ plane; 
(a) $H=0.0$, (b) $H=0.47 H_c$, 
(c) $H=0.76 H_c$ 
(d) $H=0.82 H_{c}$, (e) $H=1.01H_c$, (f) $H=1.1 H_c$, 
(g) $H=1.4 H_c$ 
(h) $H=2.35 H_{c}$
where $H_c=1.71$. 
The Chern integer for red/blue-colored 
spin-wave bands is $-1/+1$, while $0$ otherwise.}
\label{fig:a}
\end{center}
\end{figure}

\begin{figure}[tb]
\begin{center}
\includegraphics[width=94mm]{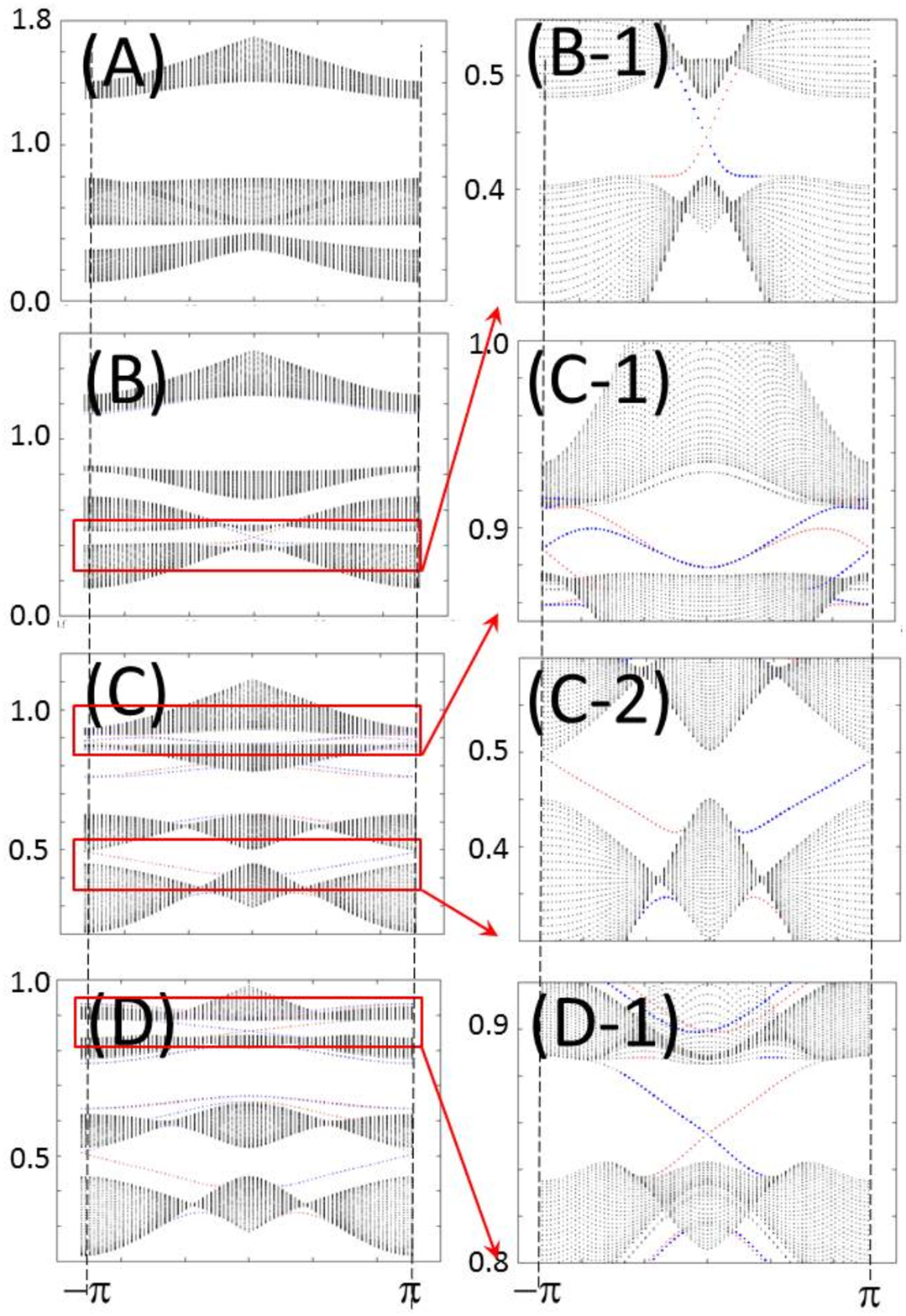}
\caption{(Color online) 
 (A-D) Wavelength-frequency dispersions calculated 
with open boundary condition along one direction ($y$-direction) 
and periodic boundary condition along the other 
(decorated square-lattice model); 
(A) $H=0.0$, (B,B-1) $H=0.47 H_{c}$, 
(C,C-1,C-2) $H=0.76 H_c$ 
(D,D-1) $H=0.82 H_c$ with $H_c=1.71$. The system 
along $y$-direction includes 40 unit cell ($L=40$). 
More than $80\%$ of  
eigenwavefunctions for red-colored points are 
localized from $y=L-3$ to $y=L$, while those    
for blue-colored points are localized from $y=1$ to $y=4$.     
Compared with Fig.~\ref{fig:a}(a-d), spectra are 
comprised also edge-mode bands, 
whose chiral dispersions runs across band gaps 
for spin wave volume modes.}
\label{fig:b}
\end{center}
\end{figure} 
       
\begin{figure}[tb]
    \includegraphics[width=94mm]{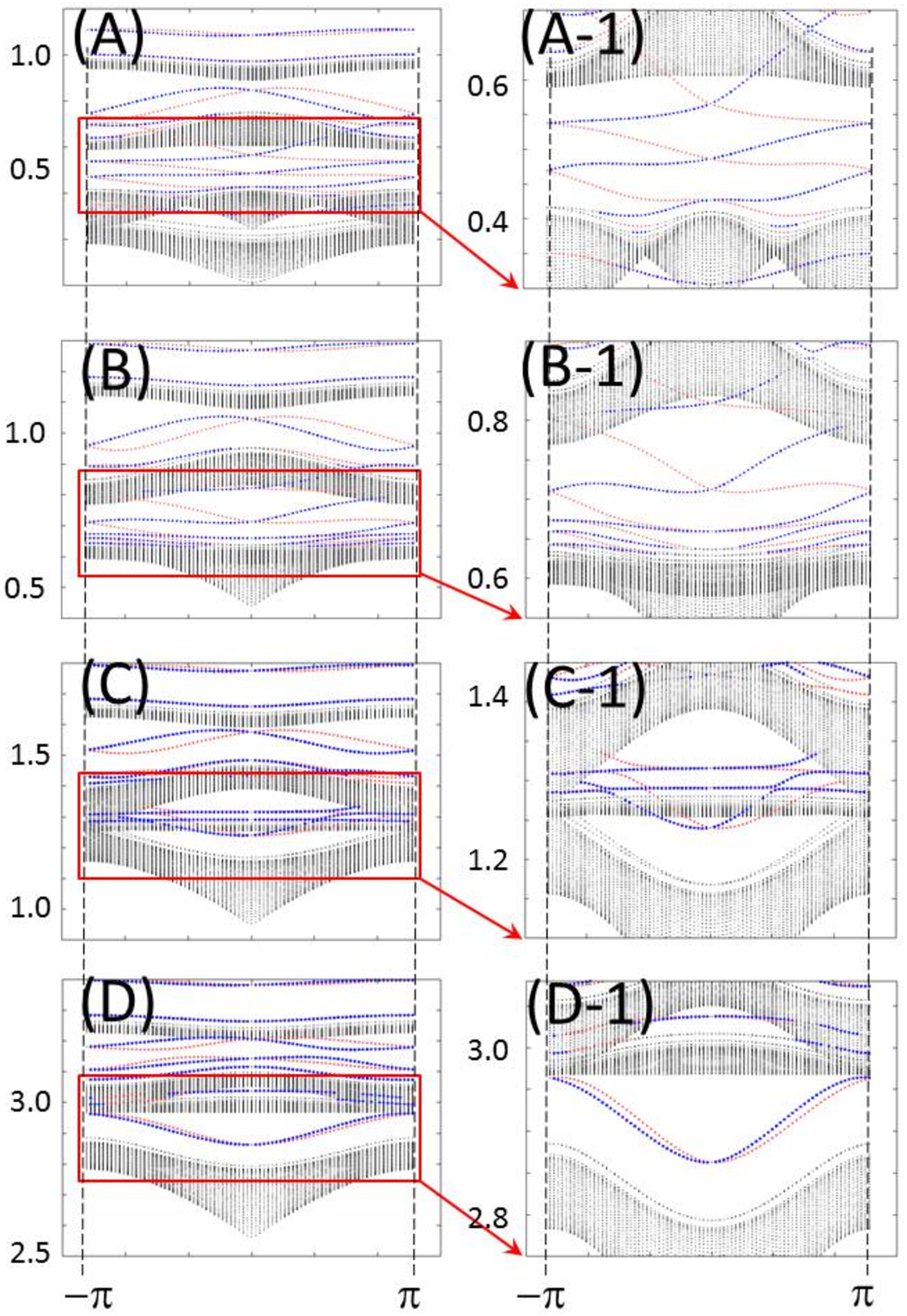}
\caption{(Color online) 
(A-D) Wavelength-frequency dispersions calculated 
with open boundary condition along one direction ($y$-direction) 
and periodic boundary condition 
along the other (decorated square-lattice model).  
(A,A-1) $H=1.01H_c$, (B,B-1) $H=1.1 H_{c}$, 
(C,C-1) $H=1.4 H_c$ 
(D,D-1) $H=2.35 H_c$ with $H_c=1.71$.}
\label{fig:c}
\end{figure}

\begin{figure}[tb]
    \includegraphics[width=90mm]{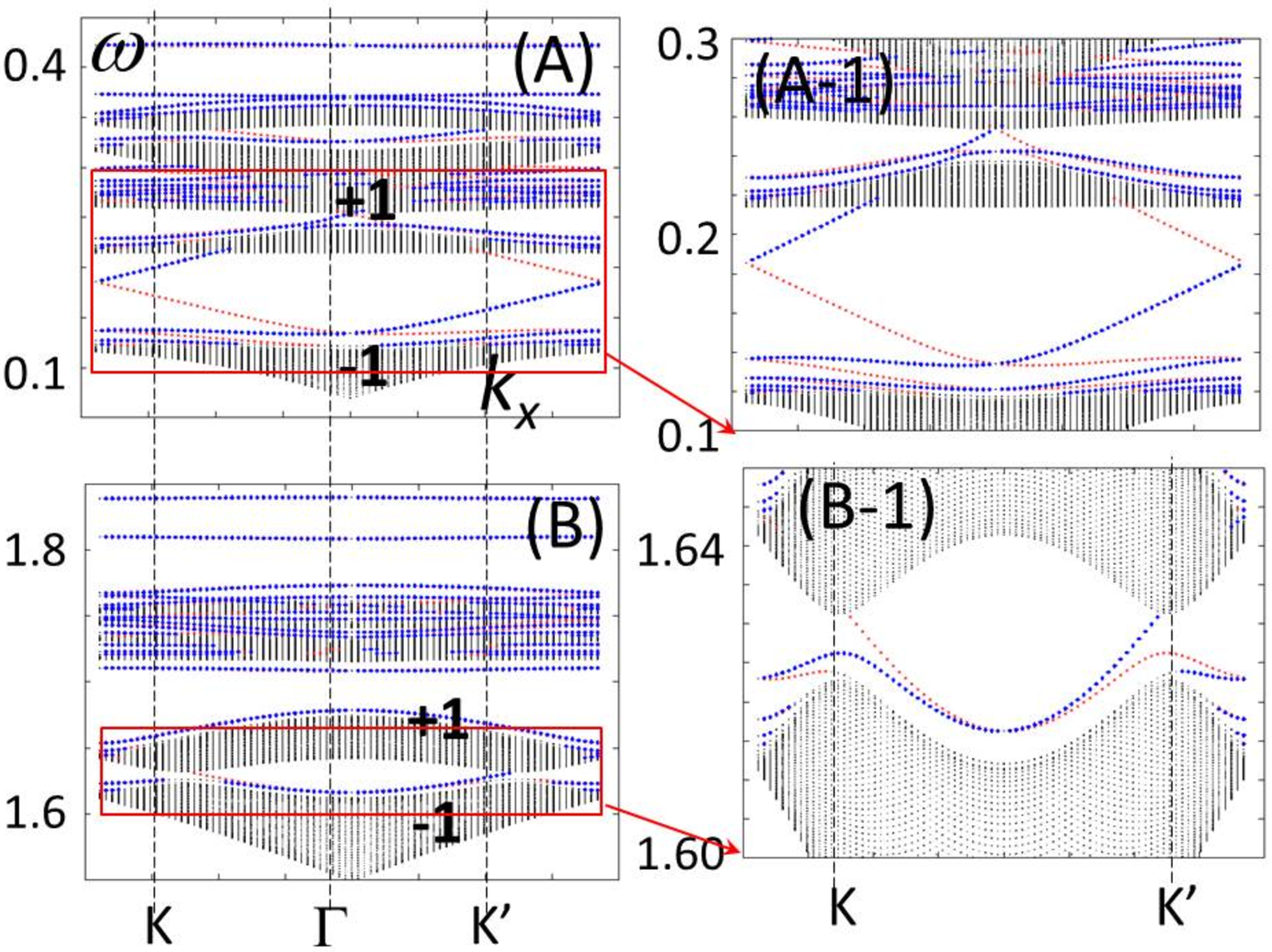}
\caption{(Color online)
(A-B) Wavelength-frequency dispersions calculated 
with open boundary condition along one direction ($y$-dir.) 
with the zigzag boundary (decorated honeycomb lattice 
model). The system along the 
$y$-direction includes 30 unit cell ($L=30$). 
Eigenwavefunctions for red-colored points are 
localized from $y=L-2$ to $y=L$ ($>80\%$), while those    
for blue-colored points are localized from $y=1$ to $y=3$.    
(A,A-1) $H=1.05 H_c$, 
(B,B-1) $H=3.5 H_c$, where $H_c=0.57$.}
\label{fig:d}
\end{figure}

\section{results}
\subsection{square-lattice model}
Results for the square-lattice model 
are summarized in Figs.~\ref{fig:a},\ref{fig:b} and \ref{fig:c}. 
Without the field, the system 
has no magnetization perpendicular to the plane, so that the 
spin-wave Hamiltonian respects both time-reversal symmetry, 
${\bm H}_{-{\bm k}}={\bm H}^{*}_{\bm k}$, and 
mirror symmetries, e.g. ${\bm H}_{(k_x,k_y)}={\bm H}_{(k_x,-k_y)}$. 
The Chern integer for all the four bands are required to be 
zero by these symmetries (Fig.~\ref{fig:a}(a)), and no chiral 
spin-wave edge modes are observed (Fig.~\ref{fig:b}(A)). 
With the field  along the $z$-direction, these symmetries 
are gone. 

On increasing the field, there appear a sequence 
of band touchings between the lowest spin-wave band and 
the second lowest one at the $\Gamma$-point 
($H=0.24 H_c$) and 
two inequivalent $X$-points ($H=0.67 H_c$). 
As a result of these band touchings, the Chern integers 
for the lowest band and the second lowest one become 
$+1$ and $-1$ respectively for 
$0.24<H/H_c<0.67$ (Fig.~\ref{fig:a}(b)), 
$-1$ and $+1$ respectively 
for $0.67<H/H_c$ (Fig.~\ref{fig:a}(c,d)). 
Correspondingly, there appears a chiral spin-wave 
edge mode propagating in the clockwise direction 
for $0.24<H/H_c<0.67$, whose dispersion runs 
across a band gap between these two spin-wave 
volume-mode bands (Fig.~\ref{fig:b}(B,B-1)). 
When the band gap closes and reopens at 
$H/H_c=0.67$, 
the chiral spin-wave edge mode changes 
its direction from clockwise to anticlockwise  
(Fig.~\ref{fig:b}(C,C-2)). The anticlockwise edge mode 
remains for a relatively larger range of the field strength, 
$0.67<H/H_c<1.4$ 
(Fig.~\ref{fig:b}(D),Fig.~\ref{fig:c}(A),(B),(A-1),(B-1)). 

There is also another sequence of band touchings  
between the third lowest spin-wave band and the highest one.
They appear at $H=0.71H_c$ 
($M$-point), $H=0.79 H_c$ 
(two inequivalent 
$X$-points) and $H=0.85 H_c$ 
($\Gamma$-point). Correspondingly, 
the first Chern integers for the third lowest band 
and the highest band become $-1$ and $+1$ 
for $0.71<H/H_c<0.79$ (Fig.~\ref{fig:a}(c)), 
$+1$ and $-1$ for $0.79<H/H_c<0.85$ (Fig.~\ref{fig:a}(d)), 
while $0$ otherwise. They lead to a  
chiral spin-wave edge mode with 
anticlockwise propagation 
($0.71<H/H_c<0.79$; Fig.~\ref{fig:b}(C,C-1)) 
and that with clockwise propagation 
($0.79<H/H_c<0.85$; Fig.~\ref{fig:b}(D,D-1)) 
between these two volume-mode bands. 

In the limit of strong field, the system becomes 
effectively time-reversal symmetric, 
${\bm H}^{*}_{\bm k}={\bm H}_{-{\bm k}}$ (consult 
also a perturbation analysis presented in sec.~IVC), 
where the Chern integers for all the four 
spin-wave volume-mode bands 
reduce to zero and the system does not support any chiral 
edge spin-wave which crosses band gaps for  
spin-wave volume-mode 
bands. Yet there still exist spin wave edge 
modes, which have parabolic dispersions at their lowest (or 
highest) frequency levels and thus consist of both 
right-moving mode and left-moving modes on a same 
side of the boundary. 
(Fig.~\ref{fig:c}(D),(D-1)). 

\subsection{honeycomb-lattice model}
Results for the decorated honeycomb lattice are 
shown in Fig.~\ref{fig:c}. Above the saturation field 
$H\ge H_{c}=0.57$, the lowest  spin-wave volume-mode 
band and the second lowest one are always 
separated by a finite band gap. The Chern integers 
for these two bands are quantized to $-1$ and $+1$ 
respectively for $H\ge H_{c}=0.57$, and  
a chiral spin-wave edge mode with the 
anticlockwise propagation cross the band 
gap between these two (Fig.~\ref{fig:d}(A,A-1)). On 
increasing the field, the band gap becomes smaller but 
always remains 
finite (Fig.~\ref{fig:d}(B,B-1)). Only in 
the strong field limit, the gap closes and 
the lowest two bands form two massless Dirac-cone 
spectra at two inequivalent 
$K$-points, ${\bm k}={\bm K}$
and ${\bm K}'$ with ${\bm K}\cdot {\bm e}_{1}
=-{\bm K}\cdot {\bm e}_2=
-{\bm K}'\cdot {\bm e}_{3}={\bm K}'\cdot {\bm e}_1 = 
\frac{2\pi}{3}$ and 
${\bm K}\cdot {\bm e}_{3}={\bm K}'\cdot {\bm e}_2=0$, 
where the Chern integers for these two bands reduce 
to zero (see also sec.IVC). In other words, the band gap 
and the chiral spin-wave edge mode which crosses over 
the gap persist even in the presence of large (but finite) 
field for the decorated honeycomb lattice 
model. \ 

\ \ 
 
\section{tight-binding descriptions}
Although we are dealing with a classical spin problem, 
the calculation procedure so far is totally 
in parallel with what standard Holstein-Primakoff 
approximation~\cite{Bloch,HP,Anderson} 
does in localized quantum spin models based on  
large $S$ expansion (where $S$ denotes 
the size of a localized quantum spin). Finding a classical 
spin configuration of a given localized spin model 
(on the order of $S^2$; treating spin as a 
classical spin) corresponds to the minimization of 
the magnetostatic energy, eq.~(\ref{magsta}) (sec.~IIIA). 
Reducing a spin model Hamiltonian into 
a quadratic form in terms of Holstein-Primakoff  
boson field (on the order of $S$) corresponds to 
linearizing the Landau-Lifshitz equation, eq.~(\ref{LL}),  
around the classical spin configuration (see sec.~IIIB). 
In fact, we diagonalized a quadratic form of the 
spin-wave Hamiltonian, eq.~(\ref{hori}), 
to obtain frequency levels of spin-wave modes 
(sec.~IV).  A tight-binding (TB) description introduced in this 
section is one approximate way of doing this 
diagonalization, which in fact gives 
useful physical pictures for 
results obtained in the previous section.   

\begin{figure}
   \includegraphics[width=70mm]{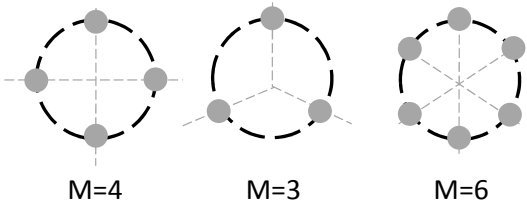}
\caption{ (Color online) A single $M$-spin cluster 
for $M=4,3,6$. $M$ spins (gray circle) align along a 
circle (radius $r$), taking an equal distance between 
their nearest neighbor spins. Each spin has a saturation   
magnetization ($M_s$) and a finite volume 
element ($\Delta V$).}
\label{fig:cluster}
\end{figure}

To construct a TB description for 
Eqs.~(\ref{eom0},\ref{gev}), let us first decompose 
the Hamiltonian defined by eq.~(\ref{hori}) 
into a diagonal part and 
off-diagonal part with respect to the $M$-spin cluster 
index;
\begin{align}
({\bm H})_{{\bm r}_i,{\bm r}_j} = ({\bm H}_{0})_{{\bm r}_i,{\bm r}_j} 
+ ({\bm H}_1)_{{\bm r}_i,{\bm r}_j}, \nn
\end{align} 
with 
\begin{align}
({\bm H}_{0})_{{\bm r}_i,{\bm r}_j} 
& = -M_s \alpha \delta_{{\bm r}_i,{\bm r}_j} 
\left(\begin{array}{cc}
1 & \\
& 1\\
\end{array}\right) \nn \\ 
&\hspace{-1.4cm} - M_s \Delta V \!\ \delta_{[{\bm r}_i],[{\bm r}_j]} 
\eta_{{\bm r}_i,{\bm r}_j}
\left(\begin{array}{cc}
f_{++}({\bm r}_i,{\bm r}_j) & f_{+-}({\bm r}_i,{\bm r}_j) \\
f_{-+}({\bm r}_i,{\bm r}_j) & f_{--}({\bm r}_i,{\bm r}_j) \\
\end{array}\right), \label{h0} \\
({\bm H}_{1})_{{\bm r}_i,{\bm r}_j} 
& = - M_s \Delta V \!\ \eta_{[{\bm r}_i],[{\bm r}_j]} 
\left(\begin{array}{cc}
f_{++}({\bm r}_i,{\bm r}_j) & f_{+-}({\bm r}_i,{\bm r}_j) \\
f_{-+}({\bm r}_i,{\bm r}_j) & f_{--}({\bm r}_i,{\bm r}_j) \\
\end{array}\right), \label{h1}
\end{align}
where $\eta_{{\bm r}_i,{\bm r}_j}=1-\delta_{{\bm r}_i,{\bm r}_j}$,   
$\eta_{[{\bm r}_i],[{\bm r}_j]}\equiv 1 - \delta_{[{\bm r}_i],[{\bm r}_j]}$ 
and $[{\bm r}_i]$ specifies a $M$-spin cluster in which the spin site 
${\bm r}_i$ are included; the 4-spin cluster for the decorated 
square lattice case ($M=4$) and the 3-spin clusters for the 
honeycomb lattice case ($M=3$); see Fig.~\ref{fig:cluster}.   

Suppose that ${\bm H}_0$ is diagonalized 
in terms of appropriate `atomic orbitals' 
localized at each $M$-spins cluster;
\begin{align}
{\bm H}_0 {\bm T}_0 = {\bm \sigma}_3 {\bm T}_0 
{\bm \sigma}_3 \tilde{\bm H}_0. \nn
\end{align}
$\tilde{\bm H}_0$ is a diagonal matrix, whose elements 
correspond to respective `atomic-orbital' levels;
The column of ${\bm T}_0$ are specified by atomic-orbital index 
($j,m$), cluster index ($n,n'$) and particle-hole 
index ($\mu$,$\nu$). The orthonormal 
condition for the new basis is given as 
${\bm T}^{\dagger}_0 {\bm \sigma}_3 {\bm T}_0 
= {\bm \sigma}_3$. In terms of the basis,  
the original eigenvalue problem, Eq.~(\ref{para}), 
reduces to  
\begin{align}
(\tilde{\bm H}_0 + \tilde{\bm H}_1 ) 
{\bm S} = {\bm \sigma}_3 {\bm S} \overline{\bm E}, \label{tb}
\end{align}
with ${\bm T} \equiv {\bm T}_0 {\bm S}$. The 
row of ${\bm S}$ and the row and the column of 
$\tilde{\bm H}_1$ are specified by atomic orbital ($j$,$m$), 
cluster ($n$,$n'$) and particle-hole $(\mu,\nu)$ indices;
\begin{align}
\big(\tilde{\bm H}_1\big)_{(j,n,\mu|m,n',\nu)} &\equiv   
\big({\bm T}^{\dagger}_0 {\bm H}_1  {\bm T}_0
\big)_{(j,n,\mu|m,n',\nu)}, \label{h1t} \\ 
\big(\tilde{\bm H}_0\big)_{(j,n,\mu|m,n',\nu)} 
&\equiv \delta_{j,m}\delta_{n,n'}\delta_{\mu,\nu} E_{0,j}, \label{h0t}
\end{align} 
where `atomic-orbital' levels $E_{0,j}$ being 
positive definite, $E_{0,j}>0$. By its construction, $\tilde{\bm H}_1$ 
has a finite matrix element only between atomic orbitals localized 
at different clusters, which thus stands for inter- or 
intra-orbital hopping terms between clusters. In terms of the 
creation/annihilation fields for the $j$-th atomic orbital localized at 
the $n$-th cluster, 
$\gamma^{\dagger}_{j,n}$/$\gamma_{j,n}$,  
Eq.~(\ref{tb}) takes the form,  
\begin{widetext}
\begin{align}
&E_{0,j} \gamma_{j,n} + \sum_{m} \sum_{n'} 
\big({\bm t}^{(+,+)}_{(j,n|m,n')} \gamma_{m,n'} 
+ {\bm t}^{(+,-)}_{(j,n|m,n')} \gamma^{\dagger}_{m,n'} \big) 
= \gamma_{j,n} \overline{E} \label{tb1} \\
&E_{0,j} \gamma^{\dagger}_{j,n} + \sum_{m} \sum_{n'} 
\big({\bm t}^{(-,+)}_{(j,n|m,n')} \gamma_{m,n'} 
+ {\bm t}^{(-,-)}_{(j,n|m,n')} \gamma^{\dagger}_{m,n'} \big) 
= - \gamma^{\dagger}_{j,n} \overline{E}. \label{tb2}
\end{align}
\end{widetext}
with 
\begin{align}
{\bm t}^{(\mu,\nu)}_{(j,n|m,n')} \equiv 
\big({\bm T}^{\dagger}_0 {\bm H}_1  {\bm T}_0
\big)_{(j,n,\mu|m,n',\nu)}. \label{transfer}
\end{align}

\begin{figure}[tb]
\begin{center}
   \includegraphics[width=68mm]{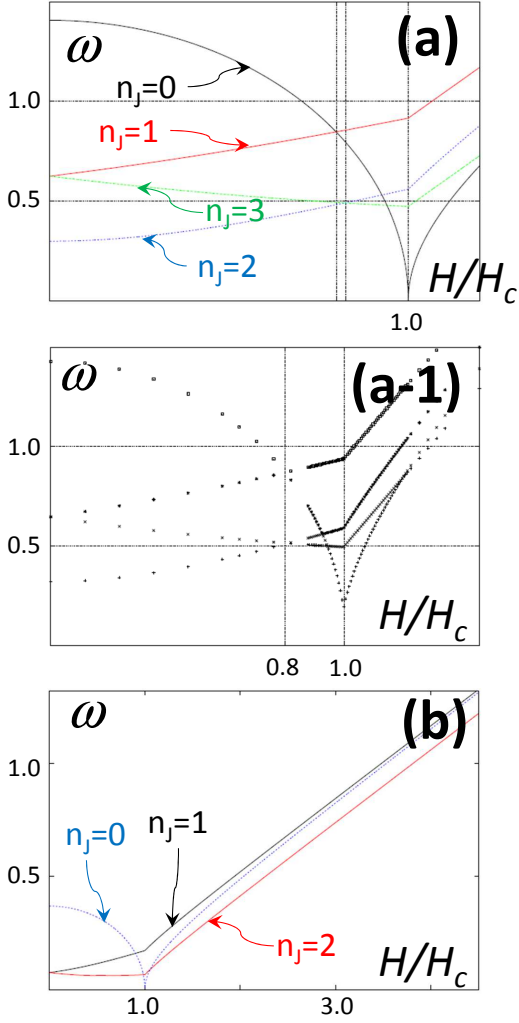}
\caption{ (Color online) (a) `Atomic-orbital' 
levels as a function 
of the field  in a single $4$-spin cluster case. 
Eq.~(\ref{h0}) is diagonalized, where the demagnetization 
field for each site ($\alpha$) 
comes from the spins in the same cluster.  
The saturation field at which $n_J=0$ becomes zero 
is around $H_c=1.26$. The level inversion between 
$n_J=1$ and $n_J=0$ is around $H/H_c=0.80$, while that 
between $n_J=3$ and $n_J=2$ is around $H/H_c=0.83$. 
(a-1) `Atomic-orbital' levels as a function 
of the field calculated from eq.~(\ref{h0}) in the 
decorated square lattice model. 
Eq.~(\ref{h0}) is diagonalized, where the demagnetization 
field for each site includes not only those from the  
spins in the same cluster but also those from spins in 
the other cluster.  
The saturation field is around 
$H_c=1.71$, where the level of $n_J=0$ goes below the others.  
(b) `Atomic-orbital' levels as a function 
of the field (single $3$-spin cluster case) with  
$H_c=0.32$.}
\label{fig:atom}
\end{center}
\end{figure}

\subsection{atomic orbitals}
To gain a useful insight on how `atomic-orbital' levels 
for ${\bm H}_0$ 
behave as a function of the out-of-plane 
field, let us first 
calculate eigenmodes for a {\it single} 
$M$-spins cluster formed by $M$ spins;  
$M$ spins align along a circle such that any neighboring two spins
are separated by a same distance (Fig.~\ref{fig:cluster}). 
As an energy minimum of the 
magnetostatic energy, the $M$ spins form 
a vortex structure with a finite out-of-plane component, 
\begin{eqnarray} 
M({\bm r}_j=r(c_{\theta_j},s_{\theta_j})) = 
(-s_{\varphi} s_{\theta_j}, s_{\varphi} c_{\theta_j},c_{\varphi})
\end{eqnarray} 
with $\theta_j=\frac{2\pi j}{M} $ ($j=1,\cdots,M$) and   
$(s_{\theta},c_{\theta})\equiv (\sin\theta,\cos\theta)$. The 
saturation field is given as $H_c/M_s \equiv 6 A_0(0) - 2 A_1(0)$ 
with 
\begin{eqnarray} 
A_0(0) \equiv \frac{\Delta V}{64\pi r^3} \sum^{M-1}_{j=1} \frac{1}{s^3_{\frac{\theta_j}{2}}}, 
\ \  
A_1(0) \equiv \frac{\Delta V}{64\pi r^3} \sum^{M-1}_{j=1} \frac{1}{s_{\frac{\theta_j}{2}}}. \nn
\end{eqnarray}
$\Delta V$ denotes a volume element of each ferromagnetic 
island (spin) and $r$ is a radius of the circle. For $H<H_c$, 
$\varphi \equiv {\rm Cos}^{-1} [-H/H_c]$ and 
$\alpha=-4A_0(0)+2A_1(0)$, while 
$\varphi=\pi$ and $\alpha=-H/M_s+2A_0(0)$ for $H>H_c$. 
Armed with these values, excitation modes are obtained 
by diagonalizing eq.~(\ref{h0}) with ${\bm r_j}=r(c_{\theta_j},s_{\theta_j})$
and $\theta_{j}\equiv \frac{2\pi j}{M}$ ($j=1,\cdots,M$). With a 
proper choice of the $U(1)$ gauge for $m_{\pm}$; 
\begin{align}
{\bm R}({\bm r}_j) = 
\left(\begin{array}{ccc}
1 & & \\
& c_{\varphi} & s_{\varphi} \\
& - s_{\varphi} & c_{\varphi} \\
\end{array}\right)  
\left(\begin{array}{ccc}
-c_{\theta_j} & - s_{\theta_j} & \\
s_{\theta_j} & -c_{\theta_j} & \\
& & 1 \\
\end{array}\right), \nn
\end{align}  
eq.~(\ref{h0}) can be readily 
diagonalized in terms of the total angular momentum $n_J$;
\begin{align}
{\bm H}_{n_J} \left(\begin{array}{c}
m_{+}(n_J) \\
m_{-}(-n_J) \\
\end{array}\right) 
= {\bm \sigma}_3 \left(\begin{array}{c}
m_{+}(n_J) \\
m_{-}(-n_J) \\
\end{array}\right) \overline{E}, \nn
\end{align} 
with 
\begin{align}
m_{\pm}(n_J) \equiv \frac{1}{\sqrt{M}} \sum^{M}_{j=1} 
e^{\pm i\frac{2\pi n_J}{M} j} m_{\pm}(\theta_j), \label{A}
\end{align} 
where $n_J$ is defined modulo $M$ $(n_J=0,1,\cdots, M-1)$.
${\bm H}_{n_J}$ takes the form of  
\begin{align}
{\bm H}_{n_J} \equiv - M_s \alpha {\bm \sigma}_0 - M_s  
\left(\begin{array}{cc}
g_{++}(n_J) & g_{+-}(n_J) \\
g_{-+}(n_J) & g_{--}(n_J) \\
\end{array}\right), \nn  
\end{align}
with 
\begin{align}
& \left(\begin{array}{cc}
g_{++}(n_J) & g_{+-}(n_J) \\
g_{-+}(n_J) & g_{--}(n_J) \\
\end{array}\right) = -2iB_0(n_J) 
c_{\varphi} {\bm \sigma}_3 \nn \\
& \hspace{0.8cm} 
+ \big\{ A_0(n_J) (-2+3c^2_{\varphi}) - A_1(n_J)(1+c^2_{\varphi}) 
\big\} {\bm \sigma}_0 \nn \\
& \hspace{1.2cm} 
- \big\{3A_0(n_J) c^2_{\varphi} 
+ A_1(n_J) (1-c^2_{\varphi})\big\} {\bm \sigma}_1, \nn 
\end{align} 
and 
\begin{align}
A_0(n_J) &\equiv \frac{\Delta V}{64 \pi r^3} 
\sum^{M-1}_{j=1} e^{iq_J j} \frac{1}{s^3_{\frac{\theta_j}{2}}}, \nn 
\end{align}
\begin{align}
A_1(n_J) &\equiv \frac{\Delta V}{64 \pi r^3} 
\sum^{M-1}_{j=1} e^{iq_J j} \frac{1}{s_{\frac{\theta_j}{2}}},  \nn 
\end{align}
\begin{align}
B_0(n_J) &\equiv  \frac{i\Delta V}{64 \pi r^3} 
\sum^{M-1}_{j=1} e^{iq_J j} 
\frac{c_{\frac{\theta_j}{2}}}{s^2_{\frac{\theta_j}{2}}}, \nn
\end{align}
with $q_J\equiv \frac{2\pi n_J}{M}$. 
Eigen-frequency with 
angular momentum $n_J$ takes the following form in 
the particle space;  
\begin{eqnarray}
\varepsilon_{n_J} = M_s\lambda_{n_J} - 2c_{\varphi} M_s B_0(n_J) \label{atl}
\end{eqnarray}
with 
\begin{align}
\lambda_{n_J} &\equiv \sqrt{\big[-\alpha 
+ 2A_0(n_J) + 2A_1(n_J)\big]} \nn \\
&\hspace{-0.7cm} \times \sqrt{\big[ 
-\alpha + 2A_0(n_J) - 6A_0(n_J) c^2_{\varphi} 
+ 2A_1(n_J) c^2_{\varphi}\big]}. \nn
\end{align}

Figs.~\ref{fig:atom}(a,b) show how the spin-wave 
excitations for a single cluster with 
$M=3$ and $M=4$, eq.~(\ref{atl}), behave as a function 
of the field respectively.  In either cases, 
doubly degenerate modes at the zero field, 
$n_J=1$ and $n_J=M-1$ ($p_{\mp}$-wave orbitals 
respectively in the 
square-lattice case; see Fig.~\ref{fig:orbital}), 
are split on increasing the field, while that with $n_J=0$ 
($s$-wave orbital) decreases its resonance frequency,  
to reach zero at the saturation field $H=H_c$. 
Above the field, the $s$-wave atomic-orbital 
level increases again, to form a quasi-degeneracy 
with the atomic-orbital level of  
$n_J=M-2$ in the large field limit;
\begin{eqnarray} 
\varepsilon_{n_J=0} = \varepsilon_{n_J=M-2} 
+ {\cal O}(1/H). \label{B}
\end{eqnarray}

\begin{figure}[tb]
   \includegraphics[width=90mm]{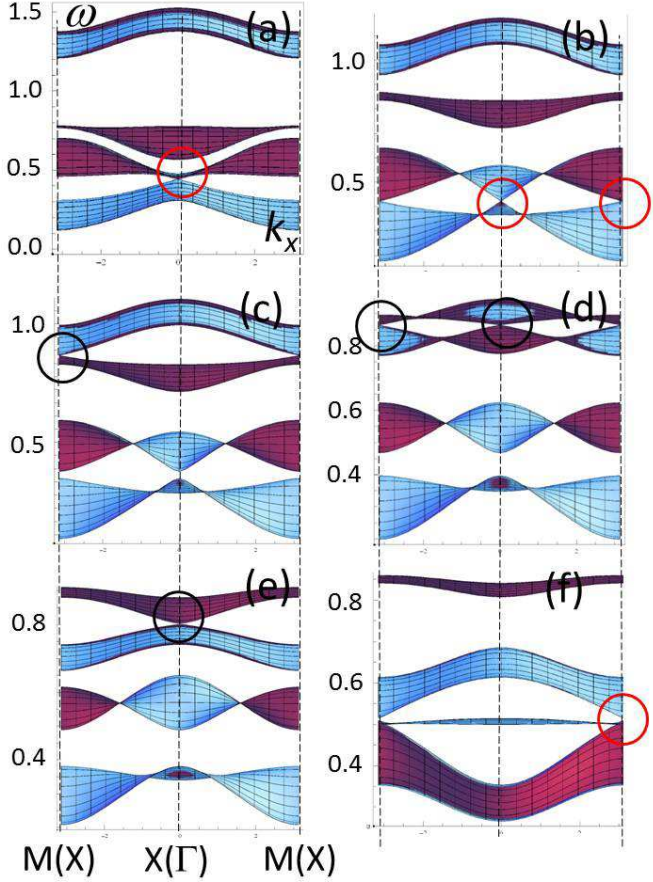}
\caption{ (Color online) Side-view of 
spin-wave band dispersions for 
decorated square-lattice model calculated from the 
tight-binding Hamiltonian, eqs.~(\ref{tb},\ref{h1t},\ref{h0t}) where 
only nearest neighbor hopping integrals are included. 
As for atomic-orbital levels, eq.~(\ref{h0t}), and the 
respective wavefunction ${\bm T}_0$ 
used in eq.~(\ref{h1t}), we used those for the single 
4-cluster. (a) $H= 0.23 H_c$, (b) $H= 0.66 H_c$, 
(c) $H= 0.71 H_c$, (d) $H= 0.79 H_c$, 
(e) $H= 0.85 H_c$, (f) $H= 1.4 H_c$,  
where $H_c$ denotes the saturation field for 
single $4$-spin cluster, ($H_c = 1.26$; 
see the caption of Fig.~\ref{fig:atom}(a)). At (a),(b) and (f), 
the lowest spin-wave band and the 2nd lowest 
one form band touchings at $\Gamma$-point, 
$X$-points, and $M$-point respectively. 
At (c),(d) and (e), 
the highest spin-wave band and the 3rd lowest  
one form band touchings at $M$-point, 
$X$-points, and $\Gamma$-point respectively.} 
\label{fig:nntb}
\end{figure}   

\begin{figure}[tb]
   \includegraphics[width=85mm]{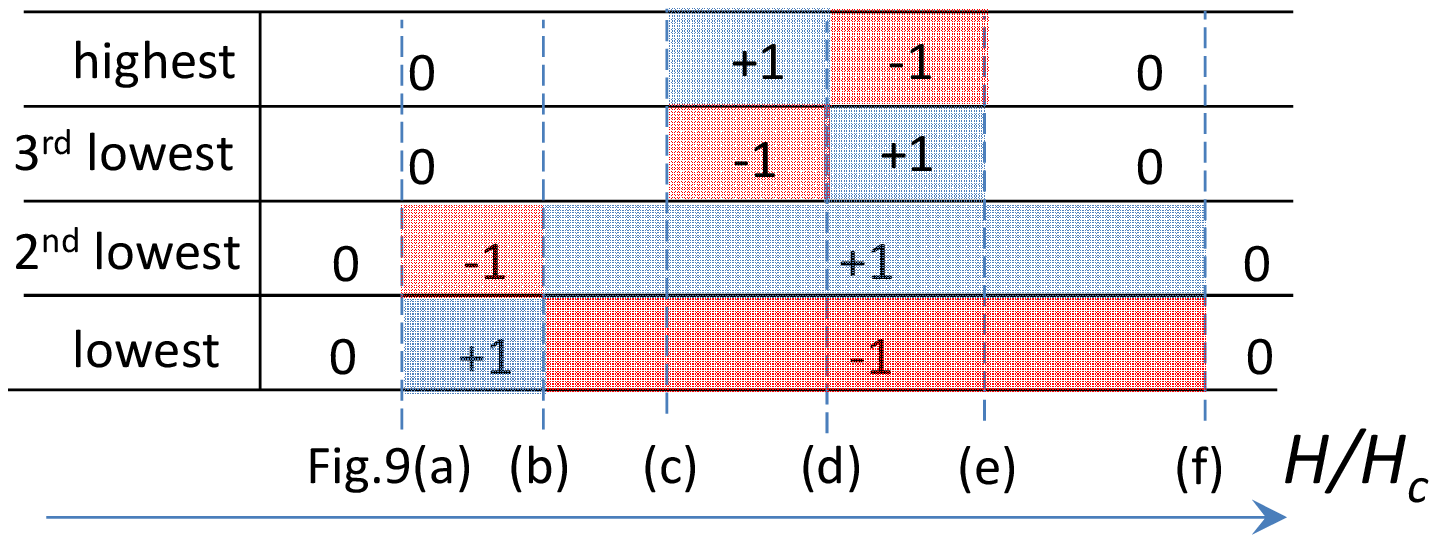}
\caption{Chern integers for four spin-wave bands 
as a function of the field. (a)-(f) depicted in the 
figure correspond  to 
the fields at which band touchings occur 
as shown in Fig.~\ref{fig:nntb}(a-f) respectively.
Note also that the third lowest spin wave band at 
Fig.~\ref{fig:nntb}(e) is mainly composed 
of $n_J=0$, while the  
2nd lowest and the lowest bands are mainly 
composed of $n_J=2,3$. From (e) to (f), 
the former band goes below the 
latter two until $H=H_c$, while, for $H>H_c$.  
it increases its frequency again, in the 
same way as the $s$-wave atomic orbital 
does in Fig.~\ref{fig:atom}(a). 
For clarity, we call the latter two as 
`2nd lowest' and `lowest', even though 
they are not during (e) - (f).}
\label{fig:nntb2}
\end{figure}   
The atomic-orbital levels shown in Figs.~\ref{fig:atom}(a,b) 
are those for a {\it single} $4$ $(3)$-spin cluster, 
where the demagnetization field stems  
only from those spins in the same cluster. 
Even when demagnetization fields from surrounding 
clusters are included, which is the case with eq.~(\ref{h0}), 
the field-dependence of the atomic-orbital levels behaves 
qualitatively in the same way as in Figs.~\ref{fig:atom}(a,b). 
Namely, the decorated square lattice model 
and honeycomb lattice model 
respects the same $4$ $(3)$-fold rotational 
symmetry as in the single $4$ $(3)$-spins cluster, 
so that only a value of the saturation field and  
specific expression for demagnetization field will 
be modified, e.g. compare Fig.~\ref{fig:atom}(a) 
with Fig.~\ref{fig:atom}(a-1).

From Figs.~\ref{fig:atom}(a,a-1), notice that there appear  
a couple of level inversions between different 
atomic orbitals, such as the one between $n_J=0$ and 
$n_J=1$ around $H=0.80H_c$, and the one  
between $n_J=2$ and $n_J=3$ around $H=0.83H_c$. 
Now that one of these two atomic orbitals 
is always either $p_x+ip_y$ or $p_x-ip_y$-like 
orbital while the others are parity even, it is expected  
from its electronic analogue~\cite{BHZ}  that 
these level inversions endow the spin-wave 
volume-mode bands 
constructed from these atomic 
orbitals with non-zero Chern integers.     
In fact, a similar type of the band inversion plays a vital 
role in the emergence of nontrivial topological phases in 
quantum spin Hall insulators.~\cite{BHZ,FK}

\subsection{TB model for the square lattice case}
To clarify how the level inversion between atomic orbitals 
leads to spin-wave bands with non-zero Chern integers, we next  
construct from Eqs.~(\ref{tb},\ref{h1t},\ref{h0t}) 
a tight-binding (TB) model for the decorated square 
lattice. ${\bm H}_0$ and corresponding ${\bm T}_0$ in 
Eqs.~(\ref{h1t},\ref{h0t}) 
are replaced by those for the single 
$4$-spin cluster. As for 
$\tilde{\bm H}_1$, only the nearest 
neighbor hopping integrals are included. Such approximations 
may be justified, because the dipolar interaction decays 
as $1/R^3$ with $R$ being a distance between two spins; 
an amplitude of the next nearest neighbor 
hopping is roughly $2\sqrt2$, $8$, and $5\sqrt{5}$ times 
larger than those of the 2nd, 3rd and 4th nearest neighbor 
hopping integrals respectively. In fact, band dispersions 
for spin-wave volume-mode bands obtained 
from this short-ranged TB model show qualitatively good 
agreements with those shown in the previous section 
(compare Figs.~\ref{fig:nntb} 
with Figs.~\ref{fig:a}). For example, all the sequences 
of the band touchings described in sec. III 
are identified near similar field 
strengths, when scaled by the respective saturation fields; 
Fig.~\ref{fig:nntb} (a-f). Correspondingly, the Chern integers 
for all the four spin-wave bands take the same 
sequence of the integer values as found in the previous 
section; Fig.~\ref{fig:nntb2}. The comparison also suggests 
that the non-parabolic characters of a certain band dispersion 
around the $\Gamma$-point in Fig.~\ref{fig:a} 
stems from long-range hopping 
integrals in $\tilde{\bm H}_1$, i.e. 
long-range character of the magnetic 
dipolar interaction, which is consistent 
with the similar feature of the 
forward volume modes.~\cite{Damon2}

The sequence of band touchings between 
the highest and the third lowest spin wave band 
results from the level inversion between the 
atomic orbital with $n_J=1$ and that with 
$n_J=0$; Fig.~\ref{fig:atom}(a),  
while the other sequence between the lowest and 
2nd lowest spin-wave band is from 
those with $n_J=2$ and $n_J=3$; Fig.~\ref{fig:atom}(a). 

\begin{figure}[tb]
\begin{center}
   \includegraphics[width=82mm]{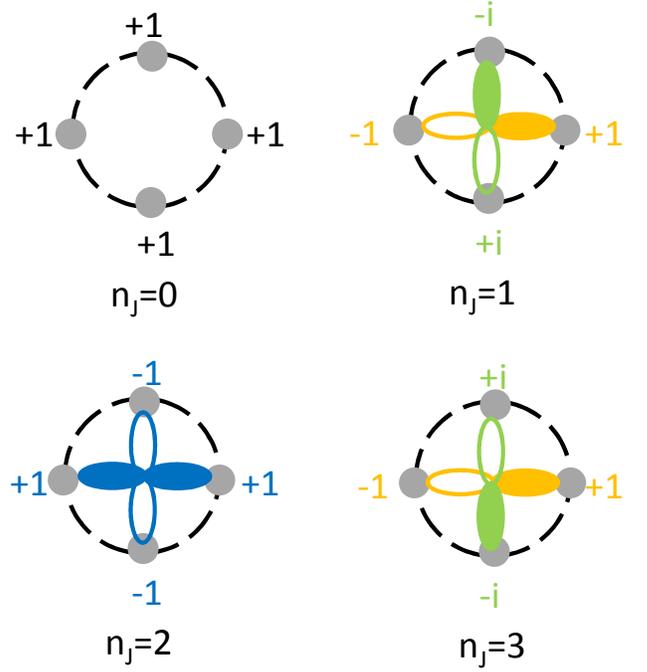}
\caption{ (Color online) Shapes of the atomic orbitals 
in the decorated square lattice model. that of $n_J=0$ 
is $s$-wave like, while those of $n_J=1,2,3$ 
are $p_{x}-ip_y$-wave, $d_{x^2-y^2}$-wave, 
$p_{x}+ip_y$-wave like respectively. }
\label{fig:orbital}
\end{center}
\end{figure}

To see this, notice first that 
the atomic orbitals with $n_J=0,1,2,3$ 
takes $s$-wave, $p_{-}\equiv p_x-ip_y$, 
$d_{x^2-y^2}$, and $p_{+}\equiv p_x+ip_y$-wave 
orbitals respectively; Fig.~\ref{fig:orbital}. 
Namely, Eq.~(\ref{A}) suggests that `atomic-orbital'   
wavefunctions for $n_J=1$ and $n_J=3$ take  
imaginary values (`$i$') along the $y$-direction, while 
take real values along the $x$-direction. Meanwhile,   
those for $n_J=0$ and $n_J=2$ always take real values; 
$n_J=0$ takes $+1$ for $x$-link and $y$-link while 
$n_J=1$ takes $+1$ and $-1$ for $x$-link and $y$-link 
respectively.  As a result, the nearest 
neighbor inter-orbital hopping integral  
between $n_J=1$ and $n_J=0$ and 
that between $n_J=2$ and 
$n_J=3$ always become pure imaginary  
along the $y$-link. 
In fact, using symmetry arguments, 
one can generally derive from eqs.~(\ref{tb1},\ref{tb2})  
a nearest-neighbor hopping model 
composed of $n_{J}=0$ and $n_J=1$ as;
\begin{align} 
&\hat{H}_{01} =  \sum_n \big( 
\varepsilon_{0} \gamma^{\dagger}_{0,n} \gamma_{0,n} 
+ \varepsilon_1 \gamma^{\dagger}_{1,n} \gamma_{1,n} \big)  \nn \\
&\ \ + \sum_{n} \sum_{\mu=x,y} \sum_{\sigma=\pm} \big(
a_{00} \gamma^{\dagger}_{0,n} \gamma_{0,n+\sigma e_{\mu}} 
+ a_{11} \gamma^{\dagger}_{1,n} \gamma_{1,n+\sigma e_{\mu}} \big) \nn \\
& \hspace{1.0cm}
+ \sum_{n}\sum_{\sigma=\pm} \big( \sigma b_{01}  
\gamma^{\dagger}_{0,n} \gamma_{1,n+\sigma e_x} 
+ {\rm h.c.} \big) \nn \\
& \hspace{1.5cm} +  \sum_{n}\sum_{\sigma=\pm} 
\big( - i \sigma b_{01} \gamma^{\dagger}_{0,n} \gamma_{1,n+\sigma e_y} 
+ {\rm h.c.} \big), \label{tb12}
\end{align}  
with real valued $a_{00}$, $a_{11}$ and $b_{01}$. 
We have ignored (or integrated out within the 
2nd order perturbation) those hopping terms which 
involve $n_J=2$ and $n_J=3$ and those 
between the particle space and the hole space. 
Such approximations 
are justified, when the atomic-orbital level of $n_J=0$ 
and that of $n_J=1$ are proximate to each other 
and when it comes to those spin-wave bands  
near these levels.

The highest and the third lowest spin-wave bands 
around $H/H_c=0.78 \sim 0.82$ are mainly composed 
of the atomic orbitals with $n_J=0$ and  
$n_J=1$ (compare Fig.~\ref{fig:atom}(a) with 
Figs.~\ref{fig:nntb}(c,d,e)) and 
thus can be approximately obtained 
from $\hat{H}_{01}$. The 
Hamiltonian in the momentum space takes 
the form~\cite{BHZ,FK} 
\begin{eqnarray}
{\bm H}_{01,{\bm k}} = \left(\begin{array}{cc}
\varepsilon_{0} + 2a_{00}(c_{k_x}+c_{k_y}) & - 2ib_{01} (s_{k_x}-is_{k_y}) \\
2ib_{01} (s_{k_x}+is_{k_y}) & \varepsilon_{1} + 2a_{11} (c_{k_x}+c_{k_y}) \\
\end{array}\right), \label{h12k}
\end{eqnarray}
with $(c_{k_x},s_{k_x})=(\cos k_x,\sin k_x)$, 
which gives momentum-frequency dispersions for 
the highest and the second highest spin-wave bands as  
\begin{align}
{\cal E}_{0,{\bm k}} &\equiv \frac{\varepsilon_{0}+\varepsilon_1}{2} 
+(a_{00}+a_{11})(c_{k_x}+c_{k_y}) 
+ \frac{\Delta_{\bm k}}{2}, \label{e1} \\
{\cal E}_{1,{\bm k}} &\equiv \frac{\varepsilon_{0}+\varepsilon_1}{2} 
+ (a_{00}+a_{11})(c_{k_x}+c_{k_y}) 
- \frac{\Delta_{\bm k}}{2}, \label{e2}
\end{align}
with
\begin{align}
&\Delta_{\bm k} \equiv \nn \\
&\sqrt{\big[\varepsilon_0-\varepsilon_1 
+ 2(a_{00}-a_{11})(c_{k_x}+c_{k_y})\big]^2 
+ 16b^2_{01} \big(s^2_{k_x} + s^2_{k_y}\big)}
\end{align}

The atomic-orbital level with $n_J=0$ and that with $n_J=1$ 
are inverted around $H=0.80 H_c$, where  
$\varepsilon_0-\varepsilon_1$ changes 
its sign from positive to negative; Fig.~\ref{fig:atom}(a).  
From their orbital shapes, the nearest neighboring hopping 
integral between $s$-wave orbitals should be always positive 
$a_{00}>0$, while that between $p_{-}$-orbitals should 
be negative $a_{11}<0$, which leads to 
$a_{00}-a_{11}>0$. These two observations 
mean that, on increasing 
the field, the two bands given by eqs.~(\ref{e1},\ref{e2}) first 
form a massless Dirac-cone spectrum at ${\bm k}=(\pi,\pi)$ 
($M$-point) when $\varepsilon_0-\varepsilon_1=4(a_{00}-a_{11})$,  
then two massless Dirac-cone spectra at ${\bm k}=(\pi,0)$ 
and ${\bm k}=(0,\pi)$ ($X$-point) 
when $\varepsilon_0-\varepsilon_1=0$, 
and finally one massless Dirac spectrum at ${\bm k}=(0,0)$ when 
$\varepsilon_0-\varepsilon_1=-4(a_{00}-a_{11})$. The  
band touching at the $M$-point is nothing but that 
in Fig.~\ref{fig:nntb}(c), those at the $X$-points are 
those in Fig.~\ref{fig:nntb}(d), 
and that at $\Gamma$-point corresponds to 
that in Fig.~\ref{fig:nntb}(e). In fact, analytic evaluations of 
eq.~(\ref{ch}) for eq.~(\ref{h12k}) show that the Chern integer 
for the highest (third lowest) spin-wave band becomes 
$+1(-1)$ for $4(a_{00}-a_{11})>\varepsilon_0-\varepsilon_1>0$ and 
$-1(+1)$ for $0>\varepsilon_0-\varepsilon_1>-4(a_{00}-a_{11})$, 
which is consistent with Fig.~\ref{fig:nntb2}. 

Similarly, the other sequence of band touchings formed 
by the lowest and second lowest spin-wave bands is 
explained in terms of the two-band models composed 
by $n_J=3$ and $n_J=2$ atomic orbitals;
\begin{eqnarray}
{\bm H}_{32,{\bm k}} = \left(\begin{array}{cc}
\varepsilon_{3} + 2a_{33}(c_{k_x}+c_{k_y}) & - 2ib_{32} (s_{k_x}-is_{k_y}) \\
2ib_{32} (s_{k_x}+is_{k_y}) & \varepsilon_{2} + 2a_{22} (c_{k_x}+c_{k_y}) \\
\end{array}\right), \label{h32k}
\end{eqnarray} 
Note that $\varepsilon_3-\varepsilon_2$ changes its sign 
from positive to negative near $H \simeq 0.83 H_c$ 
(Fig.~\ref{fig:atom}(a)), 
while $a_{33}-a_{22}$ being always negative. This means that, on 
increasing $H$, the lowest and second lowest spin-wave bands 
around $H \simeq 0.83 H_c$ 
first form a massless Dirac-cone spectrum 
at ${\bm k}=(0,0)$ when $\varepsilon_3-\varepsilon_2=-4(a_{33}-a_{22})$,  
then two massless Dirac spectra at ${\bm k}=(\pi,0)$ 
and ${\bm k}=(0,\pi)$ when $\varepsilon_3-\varepsilon_2=0$, 
and finally one massless Dirac spectrum at ${\bm k}=(\pi,\pi)$ when 
$\varepsilon_3-\varepsilon_2=4(a_{33}-a_{22})$. The  
band touching at the $\Gamma$-point is nothing but that 
in Fig.~\ref{fig:nntb}(a), those at the $X$-points are those 
in Fig.~\ref{fig:nntb}(b), 
and that at $M$-point corresponds to 
that in Fig.~\ref{fig:nntb}(f). Noting that $b_{32}$ has the same sign 
as $b_{01}$ (see Fig.~\ref{fig:orbital}), one  
can also see from the previous evaluation that the Chern integer 
for the second lowest (lowest) spin-wave band becomes 
$-1(+1)$ for $-4(a_{33}-a_{22})>\varepsilon_3-\varepsilon_2>0$ and 
$+1(-1)$ for $0>\varepsilon_3-\varepsilon_2>4(a_{33}-a_{22})$, 
which is consistent with Fig.~\ref{fig:nntb2}.

\subsection{TB model for the honeycomb lattice case}
In the decorated honeycomb lattice 
model, we have observed in sec. III 
a finite band gap between 
the lowest spin-wave band and second lowest one, 
which are connected by a dispersion of a chiral edge 
mode. 
The gap and chiral edge mode persists for 
a sufficiently large field $H$. Based on a tight-binding 
model, we will employ a perturbation analysis 
from the large field limit and argue that the gap closes 
at two inequivalent $K$-points only in the  
limiting case ($|H|\rightarrow \infty$), 
where both the time-reversal 
symmetry and hexagonal spatial symmetry are effectively 
recovered. More accurately, we will show that an effective 
spin-wave Hamiltonian in the 
large field limit respects these two symmetries 
within the order of ${\cal O}(1)$, while it starts to break  
them from ${\cal O}(1/H)$. 
As a result, within the order of ${\cal O}(1)$, 
the lowest and second lowest spin-wave band compose  
massless Dirac spectra at the $K$-points. 
Once the ${\cal O}(1/H)$-corrections are 
included, the time-reversal symmetry is broken and 
the hexagonal symmetry ($C_{6v}$) 
reduces to its subgroup symmetry ($C_6$), 
which leads to a finite band gap at the $K$-points. 
These symmetry breakings also endow the two bands 
with a non-zero Chern integer with opposite signs, 
which results in the emergence 
of chiral edge mode within the band gap. 

The perturbation analysis begins 
with a tight-binding Hamiltonian for the honeycomb 
lattice model, eq.~(\ref{tb});
\begin{align}
\tilde{\bm H}_0 + \tilde{\bm H}_1 
= H {\bm \sigma}_0 + \lambda {\bm V}_1 + \lambda {\bm V}_2 
\end{align} 
where ${\bm \sigma}_0$ 
is a $2$ by $2$ unit matrix in the particle-hole space
and both ${\bm V}_1$ and ${\bm V}_2$ are 
on the order of ${\cal O}(1)$.  For a bookkeeping, 
we put $\lambda$, which can be set to 1 
from the outset [those terms with $\lambda$ 
are ${\cal O}(1)$, those with $\lambda^2$ 
are ${\cal O}(1/H)$ and those with $\lambda^3$ are 
${\cal O}(1/H^2)$; see below]. ${\bm V}_1$ consists of 
on-cluster `atomic-orbital' levels and  hopping terms in 
the excitonic channel, 
while ${\bm V}_2$ consists only of those in the 
Cooper channel;
\begin{align}
{\bm V}_1 &\equiv \tilde{\bm H}_0
- H {\bm \sigma}_0 + 
\left(\begin{array}{cc}
{\bm t}^{(+,+)} & {\bm 0} \\
{\bm 0} & {\bm t}^{(-,-)} \\
\end{array}\right), \nn \\
{\bm V}_2 &\equiv 
\left(\begin{array}{cc}
{\bm 0} & {\bm t}^{(+,-)} \\
{\bm t}^{(-,+)} &  {\bm 0} \\
\end{array}\right), \nn  
\end{align}    
In the large field limit, all the spin-wave excitations 
reduce to the ferromagnetic resonance (FMR) 
with its resonance frequency being $H$. 
Once the 
${\cal O}(1)$-corrections (${\bm V}_1$,${\bm V}_2$) are 
included, the FMR resonance is expected to split into a 
couple of spin-wave bands whose bandwidth 
are at most on the order of unity. To see this 
situation,  let us erase those terms in the Cooper 
channel within a given order accuracy in $1/H$, and 
derive an effective Hamiltonian only for the particle space. 
 
With a matrix satisfying 
${\bm \sigma}_3 {\bm \rho}{\bm \sigma}_3 = {\bm \rho}^{\dagger}$, 
the transformed Hamiltonian takes the form 
\begin{align}
{\bm H}_{\rm eff} 
&\equiv e^{-i\lambda {\bm \sigma}_3 {\bm \rho} {\bm \sigma}_3} \!\ 
(H  + \lambda {\bm V}_1 
+ \lambda {\bm V}_2) \!\ e^{i\lambda{\bm \rho}} \nn \\
& = (1-i \lambda {\bm \sigma}_3 {\bm \rho} {\bm \sigma}_3 
- \frac{\lambda^2}{2} {\bm \sigma}_3  {\bm \rho}^2 {\bm \sigma}_3 + \cdots ) 
\nn \\
&\ \  \times (H + \lambda {\bm V}_1 
+ \lambda {\bm V}_2) 
(1+ i \lambda {\bm \rho}  
- \frac{\lambda^2}{2}  {\bm \rho}^2 + \cdots ) \nn \\
&= H  + \lambda {\bm V}_1 
+ \lambda {\bm V}_2 - \lambda 
i{\bm \sigma}_3  {\bm \rho} {\bm \sigma}_3 
H  + i \lambda H   {\bm \rho} \nn \\
& \ \ - \frac{\lambda^2}{2} {\bm \sigma}_3  {\bm \rho}^2 {\bm \sigma}_3 
H  - \frac{\lambda^2}{2} H {\bm \rho}^2 
+ \lambda^2 {\bm \sigma}_3  {\bm \rho} {\bm \sigma}_3 
H  {\bm \rho} \nn \\
& \hspace{0.5cm} 
 - i \lambda^2 {\bm \sigma}_3  {\bm \rho} {\bm \sigma}_3 {\bm V}_1 
+ i \lambda^2 {\bm V}_1  {\bm \rho} - i\lambda^2 
{\bm \sigma}_3  {\bm \rho} {\bm \sigma}_3 
{\bm V}_2 \nn \\ 
&\ \ \ \ \  + i  \lambda^2 {\bm V}_2  {\bm \rho} + {\cal O}(\lambda^3)  
\label{rot1}   
\end{align}
We choose $\rho$ such that all the 
matrix elements in the Cooper channels will cancel  
each other within the order of ${\cal O}(1)$;
\begin{align} 
{\bm V}_2 = i {\bm \sigma}_3 {\bm \rho} {\bm \sigma}_3 H  
- i H {\bm \rho}. \nn  
\end{align}  
Or equivalently, 
\begin{align}
\big({\bm \rho}\big)_{n,\overline{m}} 
= i \frac{\big({\bm V}_2\big)_{n,\overline{m}}}{2H}, \ \ \ 
\big({\bm \rho}\big)_{\overline{n},m} 
= i \frac{\big({\bm V}_2\big)_{\overline{n},m}}{2H}, \label{gene}
\end{align}
where $\overline{n}$ is for the indices in the hole space 
and $n$ is for those in the particle space; 
$({\bm \sigma}_3)_{n,m} = \delta_{n,m}$ 
and $({\bm \sigma}_3)_{\overline{n},\overline{m}} = -\delta_{n,m}$. 
With this rotated frame, all the matrix elements in  
the Cooper channel are at most 
on the order of ${\cal O}(1/H)$;
\begin{align}
{\bm H}_{\rm eff}  
& = H + \lambda{\bm V}_1 - \frac{i \lambda^2}{2} 
\big({\bm \sigma}_3 {\bm \rho}{\bm \sigma}_3 {\bm V}_2 
- {\bm V}_2 {\bm \rho} \big) \nn \\
& \ \ - i \lambda^2 
\big( {\bm \sigma}_3 {\bm \rho}{\bm \sigma}_3 {\bm V}_1 
- {\bm V}_1 {\bm \rho} \big). \label{rot2}
\end{align}
The last two terms have matrix elements 
in Cooper channels. 
When we further rotate in the particle-hole 
space such that they will be set off by generated terms,  
these two terms simply result in higher order terms, 
${\cal O}(1/H^2)$, while the remaining terms being 
kept intact. 
We thus drop them by hand, to keep the first four terms 
as the effective Hamiltonian. 
On the substitution of eq.~(\ref{gene}) into 
eq.~(\ref{rot2}), we then have the effective Hamiltonian to 
the order of $1/H$ as  
\begin{align}
&\big({\bm H}^{(2)}_{\rm eff}\big)_{n,m} 
= H \delta_{n,m}  + \big({\bm V}_1\big)_{n,m} \nn \\
& \hspace{2.5cm}
+ \frac{1}{2H} \sum_{\overline{p}} \big({\bm V}_2\big)_{n,\overline{p}}
\big({\bm V}_2\big)_{\overline{p},m}. \label{rot3}
\end{align}  
The superscript `$(2)$' in the left hand side denotes that 
the effective Hamiltonian is asymptotically exact  
within the 2nd order in $\lambda$ (or within the first 
order in $1/H$). The sum with respect to 
$\overline{p}$ is taken only 
over the hole space. From eq.~(\ref{rot3}), 
one can ready see that, once the 
${\cal O}(1)$-corrections 
(${\bm V}_1$) are included, the FMR resonance localized 
at $H$ is split into a couple of spin-wave bands 
whose bandwidth are at most on the order of unity.

Within the order of ${\cal O}(1)$, the effective 
Hamiltonian derived so far 
is invariant under the time-reversal operation and 
hexagonal symmetry operations. To see this, let us 
focus on the first two terms of eq.~(\ref{rot3}). 
With the atomic orbital index ($n_J=j,m$) 
and cluster index ($n$,$n'$) being made explicit, they take 
the following form  
\begin{align}
\big({\bm H}^{(1)}_{\rm eff}\big)_{(j,n|m,n')} 
& = \delta_{n,n'} \delta_{j,m} \varepsilon_{j} + 
{\bm t}^{(+,+)}_{(j,n|m,n')}. \nn 
\end{align}
In the leading 
order in $1/H$, the inter/intra-orbital hopping 
integral between 
an orbital with $n_J=j$ at the $n$-th cluster 
and that with $n_J=m$ at the $n'$-th 
cluster is given by   
\begin{align}
{\bm t}^{(+,+)}_{(j,n|m,n')} 
= \sum_{\theta_l,\theta_{l'}} 
e^{i(j+1)\theta_l-i(m+1)\theta_{l'}} \frac{1}{6R^3}. \label{tpp}
\end{align}
$\theta_{l^{(\prime)}}$ $(l^{(\prime)}=1,2,3)$ in the right hand side 
specifies a spatial location of a ferromagnetic spin within a cluster. 
Within a cluster which encompasses an A-sublattice site or 
B-sublattice site at $(x_n,y_n)$, we take 
$\theta_{l} \equiv \frac{2\pi l}{3}$ or $\pi + 
\frac{2\pi l}{3}$ respectively, such that the location of 
the ferromagnetic spin is always  
given by $(x_n-r \sin\theta_l,y_n+\cos\theta_l)$. 
$R$ denotes a spatial distance between a 
ferromagnetic spin specified by $(\theta_l,n)$ and  
that by $(\theta_{l'},n')$;
\begin{eqnarray}
R \equiv \Bigg| \left(\begin{array}{c}
x_n - r\sin \theta_l \\
y_n + r\cos \theta_l \\
\end{array}\right) - \left(\begin{array}{c}
x_{n'} - r\sin \theta_{l'} \\
y_{n'} + r\cos \theta_{l'} \\
\end{array}\right)\Bigg|. \nn
\end{eqnarray} 

Within the order of ${\cal O}(1)$, 
a complex conjugatation of 
hopping integrals can be readily 
set off by a sign change of the {\it orbital} angular 
momentum $n_L\equiv n_J+1$. Namely, 
the complex conjugate of eq.~(\ref{tpp}) 
is transformed to itself by a proper exchange 
between $n_L=+1$ ($n_J=0$) and 
$n_L=2\equiv -1 \!\ ({\rm mod} \!\ \!\ 3)$ ($n_J=1$). 
This in combination with eq.~(\ref{B}) 
indicates that the effective Hamiltonian up to 
the order of ${\cal O}(1)$ is invariant under the 
following time-reversal operation;
\begin{align}
&\big({\bm H}^{(1)}_{\rm eff}\big)^{*}_{(j,n|m,n')}
= {\bm Q}_{jj'} 
\big({\bm H}^{(1)}_{\rm eff}\big)_{(j',n|m',n')}
{{\bm Q}^t}_{m'm}. \label{time-reversal} 
\end{align}
with a proper basis change 
\begin{eqnarray}
{\bm Q} \equiv \left(\begin{array}{ccc}
 & 1 & \\ 
1 & & \\
& & 1 \\
\end{array}\right), \nn 
\end{eqnarray}
which exchanges $n_L=+1$ $(n_J=0)$ and 
$n_L=-1$ $(n_J=1)$, 
while keeps $n_L=0$ ($n_J=2$) intact.  
It is also invariant under three mirror operations  
in the hexagonal symmetry, $\sigma_{v,1}$, 
$\sigma_{v,2}$, $\sigma_{v,3}$, $\pi$-rotation 
$C_2$ and $\frac{2\pi}{3}$-rotation $C_3$; 
\begin{align}
\big({\bm H}^{(1)}_{\rm eff}
\big)_{(j,{\sigma}_{v,1}(n)|m,\sigma_{v,1}(n'))} 
& = {\bm Q}_{jj'} 
\big({\bm H}^{(1)}_{\rm eff}\big)_{(j',n|m',n')}
{{\bm Q}^t}_{m'm}, \nn \\
%
%
\big({\bm H}^{(1)}_{\rm eff}
\big)_{(j,\sigma_{v,2}(n)|m,\sigma_{v,2}(n'))} 
& = {\bm O}_{jj'}
\big({\bm H}^{(1)}_{\rm eff}\big)_{(j',n|m',n')}
{{\bm O}^{\dagger}}_{m'm}, \nn \\
\big({\bm H}^{(1)}_{\rm eff}
\big)_{(j,\sigma_{v,3}(n)|m,\sigma_{v,3}(n'))} 
& = {{\bm O}^{\dagger}}_{jj'} 
\big({\bm H}^{(1)}_{\rm eff}\big)_{(j',n|m',n')}
{\bm O}_{m'm}, \nn \\
\big({\bm H}^{(1)}_{\rm eff}\big)_{(j,C_2(n)|m,C_2(n'))} 
& =  
\big({\bm H}^{(1)}_{\rm eff}\big)_{(j',n|m',n')}, \nn \\
\big({\bm H}^{(1)}_{\rm eff}\big)_{(j,C_3(n)|m,C_3(n'))} 
& = {{\bm P}^{\dagger}}_{jj'} 
\big({\bm H}^{(1)}_{\rm eff}\big)_{(j',n|m',n')}
{\bm P}_{m'm}, \nn
\end{align}  
respectively with
\begin{eqnarray}
{\bm O} \equiv 
\left(\begin{array}{ccc}
 & e^{-i\frac{2\pi}{3}} & \\ 
e^{i\frac{2\pi}{3}} & & \\
& & 1 \\
\end{array}\right), \ 
{\bm P} \equiv 
\left(\begin{array}{ccc}
e^{i\frac{2\pi}{3}} & & \\ 
& e^{-i\frac{2\pi}{3}} & \\
& & 1 \\
\end{array}\right). \nn 
\end{eqnarray}  
$\sigma_{v,\nu}$ ($\nu=1,2,3$) denotes the mirror operation 
with respect to the plane subtended 
by ${\bm e}_{\nu}$  ($\nu=1,2,3$; see Fig.~\ref{fig:model}) 
and ${\bm e}_z$ (a unit vector 
normal to the plane), $C_2$ is the two-fold rotation 
which exchanges A-sublattice and B-sublattice and 
$C_3$ is the three-fold rotation within the plane 
(see Fig.~\ref{fig:model}(b)).   

Owing to the hexagonal symmetry, the lowest 
two spin wave bands obtained from ${\bm H}^{(1)}_{\rm eff}$ 
comprise two massless Dirac spectra at two inequivalent 
$K$-points. 
Once the ${\cal O}(1/H)$ corrections are included, 
e.g. $\varepsilon_0 \ne \varepsilon_1$, 
the time-reversal symmetry 
is lost and the hexagonal ($C_{6v}$) symmetry 
($C_2,C_3,C^{-1}_3,\sigma_{v,1},\sigma_{v,2},\sigma_{v,3},
\sigma_{d,1},\sigma_{d,2},\sigma_{d,3}$) with 
$\sigma_{d,\nu} \equiv C_2 \cdot \sigma_{v,\nu}$ 
reduces to $C_6$ symmetry $(C_2,C_3,C^{-1}_3)$. As 
a result, Dirac spectra at the 
$K$-points acquire a finite mass, 
which gives to the lowest two bands non-zero Chern integers.

\begin{figure}
\begin{center}
   \includegraphics[width=60mm]{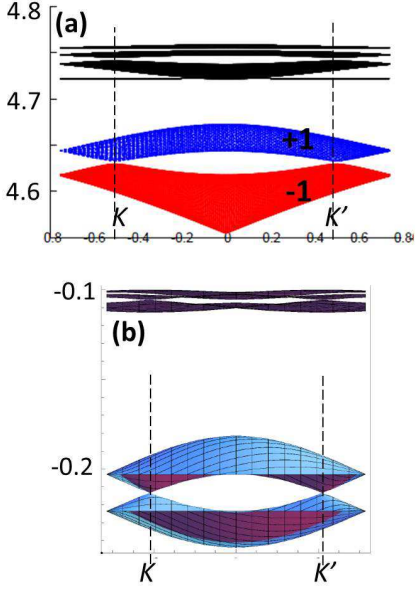}
\caption{ (Color online) (a) Side-view of spin-wave band dispersions 
for the decorated honeycomb lattice model under a sufficiently 
strong field ($H=5.0$). Because of a small but finite 
band gap at two $K$-points, the Chern integers for the 
lowest two bands are $-1$ and $+1$ respectively. 
 (b) Side-view of spin-wave band dispersions 
calculated from the effective Hamiltonian to 
the order of ${\cal O}(1)$. For the `atomic-orbital' levels, 
we use those for a single $3$-spin cluster. 
To evaluate the nearest neighbor hopping integral 
within the order of ${\cal O}(1)$, 
we use Eq.~(\ref{tpp}).   
When $H=5.0$ is added, 
the resonance frequencies of the spin-wave bands 
in Fig.~\ref{fig:unit}(b) become comparable 
to those in Fig.~\ref{fig:unit}(a).}
\label{fig:unit}
\end{center}
\end{figure}

The Chern integers for the lowest two spin-wave 
band can be evaluated from a nearest neighboring 
(NN) TB model. From the symmetry point of view, 
the NN TB Hamiltonian in the momentum space 
reads, 
\begin{align}
& {\bm H}_{{\rm NNTB},\bm k} = \nn \\
& 
\left(\begin{array}{cccccc}
\varepsilon_0 & & & \alpha_0 a_{0,{\bm k}} & 
\beta a_{1,{\bm k}}  & \gamma_0 a_{2,{\bm k}} \\
& \varepsilon_1 & & \beta a_{2,{\bm k}}  & \alpha_1 a_{0,{\bm k}} 
&\gamma_1 a_{1,{\bm k}}  \\
& & \varepsilon_2 & \gamma_0 a_{1,{\bm k}} 
& \gamma_1 a_{2,{\bm k}} & \eta a_{0,{\bm k}} \\
\alpha_0 a^{*}_{0,{\bm k}}&  \beta a^{*}_{2,{\bm k}}& 
\gamma_0 a^{*}_{1,{\bm k}} & \varepsilon_0 & & \\
\beta a^{*}_{1,{\bm k}} & \alpha_1 a^{*}_{0,{\bm k}} & 
\gamma_1 a^{*}_{2,{\bm k}} & & \varepsilon_1 & \\
\gamma_0 a^{*}_{2,{\bm k}} & 
\gamma_1 a^{*}_{1,{\bm k}} & 
\eta a^{*}_{0,{\bm k}}& & & \varepsilon_2 \\
\end{array}\right) \nn
\end{align}
with 
\begin{align}
a_{0,{\bm k}} & \equiv e^{-i{\bm k}{\bm e}_1} 
+ e^{-i{\bm k}{\bm e}_2} + e^{-i{\bm k}{\bm e}_3}, \nn \\
a_{1,{\bm k}} & \equiv e^{-i{\bm k}{\bm e}_1} 
+ e^{-i\frac{2\pi}{3}} e^{-i{\bm k}{\bm e}_2} + 
e^{i\frac{2\pi}{3}} e^{-i{\bm k}{\bm e}_3}, \nn \\
a_{2,{\bm k}} & \equiv e^{-i{\bm k}{\bm e}_1} 
+ e^{i\frac{2\pi}{3}} e^{-i{\bm k}{\bm e}_2} + 
e^{-i\frac{2\pi}{3}} e^{-i{\bm k}{\bm e}_3} , \nn
\end{align} 
The first three columns and rows are for the three 
atomic orbitals encompassing an A-sublattice site, 
while the latter three are for those encompassing 
a B-sublattice site. $\varepsilon_j$ stands for a  
level for an atomic orbital with $n_J=j$ $(j=0,1,2)$.  
$\alpha_0$, $\alpha_1$, $\beta$, $\eta$, $\gamma_0$ 
and $\gamma_1$ are NN inter or intra-orbital (effective) 
transfer integrals, which can be evaluated 
from eq.~(\ref{rot3}) up to ${\cal O}(1/H)$.   
It is clear from Eq.~(\ref{tpp}) that, 
within the order of ${\cal O}(1)$, 
$\gamma_0=\gamma_1$, 
$\alpha_0=\alpha_1$, and $\varepsilon_0=\varepsilon_1$, 
which makes the lowest two bands 
form massless Dirac spectra at the $K$-points; 
Fig.~\ref{fig:unit}(b).  A comparison between 
Fig.~\ref{fig:unit}(a) and 
Fig.~\ref{fig:unit}(b) suggests that 
the present NN TB Hamiltonian can 
qualitatively well reproduce the band structure 
of the lowest two bands in the large field limit, 
expect for a non-parabolic band structure of 
the lowest band near the $\Gamma$-point.

Once finite $\Delta\gamma \equiv  
\gamma_0-\gamma_1$, 
$\Delta \alpha \equiv \alpha_0-\alpha_1$ and 
$\Delta \varepsilon \equiv \varepsilon_0 - \varepsilon_1$ are 
included, the exchange between $n_L=+1$ and 
$n_L=-1$ changes the signs of these terms,  
so that the time-reversal symmetry is broken and 
the hexagonal symmetry reduces to the $C_6$ 
symmetry. These symmetry reductions 
give a finite mass to the Dirac spectra. 
The mass can be evaluated from  
2 by 2 Dirac Hamiltonians for the lowest two 
spin-wave bands, which can be obtained via $k\cdot p$ 
perturbation around these $K$-points;  
\begin{align}
{\bm H}^{2\times 2}_{{\bm k}={\bm K}+{\bm p}}  
&= \frac{1}{2}\Big(\Delta \varepsilon \!\ \sin^2\frac{\theta}{2} 
- 3\Delta \gamma \!\ \sin \theta \Big) 
{\bm \sigma}_3  \nn \\
&\hspace{-1cm} + \frac{3}{2} \Big(\eta \!\ \cos^2\frac{\theta}{2} 
- \beta \!\ \sin^2 \frac{\theta}{2} \Big)
(p_x {\bm \sigma}_1 - p_y {\bm \sigma}_2), \nn
\end{align}
and 
\begin{align}
{\bm H}^{2\times 2}_{{\bm k}={\bm K}^{\prime}+{\bm p}}  
&= - \frac{1}{2}\Big(\Delta \varepsilon \!\ \sin^2\frac{\theta}{2} 
- 3\Delta \gamma \!\ \sin \theta \Big) 
{\bm \sigma}_3  \nn \\
&\hspace{-1.0cm} + \frac{3}{2} \Big(\eta  \!\ \cos^2\frac{\theta}{2} 
- \beta \!\ \sin^2 \frac{\theta}{2} \Big)
(-p_x {\bm \sigma}_1 - p_y {\bm \sigma}_2), \nn
\end{align}
with 
\begin{align}
\tan \theta 
\equiv 3 \frac{\gamma_1+\gamma_2}{\varepsilon_0-\varepsilon_1}. \nn
\end{align}
From these Hamiltonians, 
the Chern integers for the lowest and second 
lowest spin-wave bands are evaluated to be $\sigma$ 
and $-\sigma$ respectively with     
\begin{align}
\sigma \equiv {\rm sign} \Big(
\Delta \varepsilon \!\ \sin^2 \frac{\theta}{2} 
- 3\Delta \gamma \!\ \sin \theta\Big). \nn
\end{align}
A substitution of actual numbers into the parameters 
in the right hand side shows that $\sigma=1$, 
which is consistent 
with previous numerical evaluations in Sec.III. 
The non-zero Chern integers for the lowest two 
spin-wave bands results in an edge  
mode with the anticlockwise propagation, which 
has a chiral dispersion between these two bands.  

\section{Micromagnetic Simulation}

\begin{figure}[tb]
\begin{center}
   \includegraphics[width=87mm]{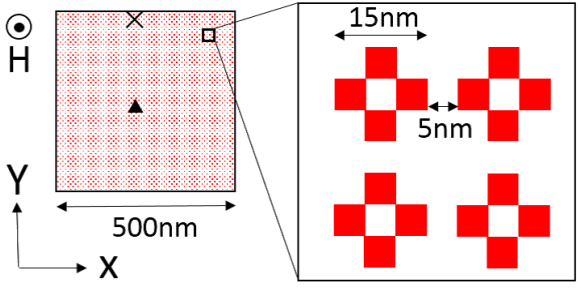}
\caption{ (Color online) Schematic view of a simulated 
system that comprises ferromagnetic nanograin. Although we took 
the size of the ferromagnetic nanograin to be 
$5\times 5 \times 5$ nm$^3$ as a demonstration, 
the present simulation is scale free.   
In the begining of the simulation, 
we apply a pulse field either at the 
center (marked by a black triangle) or around the 
boundary (marked by a black cross).}
\label{fig:system}
\end{center}
\end{figure}

To uphold the existence of proposed chiral spin-wave  
edge mode by a standard method in the field,  
we perform a micromagnetic simulation by solving numerically 
the Landau-Lifshitz-Gilbert equation for the square-lattice 
model. We calculate  
magnetization dynamics by employing the 
4th order Runge-Kutta method 
with a time step $\Delta t=1$ ps. Fig.~\ref{fig:system} 
schematically shows an entire system studied 
in the present micromagnetic simulation. 
It consists of 4 ferromagnetic nanograins in the unit cell. 
Although we took the size of the ferromagnetic nanograin to be  
$5\times 5 \times 5$ nm$^3$ as a demonstration, 
the system is scalable; 
the simulation does not include any 
short-range exchange interactions. The 
saturation magnetization and Gilbert damping coefficient 
of the ferromagnetic grain are $1.75\ $J/$\mu$m and 
$1.0 \times 10^{-5}$ respectively. 
We regard each nanograin as a uniform magnet, to 
assign single spin degree of freedom to each nanograin.  
Different ferromagnetic nanograins are 
coupled with one another through the magnetic  
dipolar interaction. The simulated system 
($0<X<L$ and $0<Y<L$; Fig.~\ref{fig:system}) includes 
$25 \times 25$ unit cells. Without the field,  
the magnetization of each grain lies within the plane 
due to the dipolar interaction. Under a large 
out-of-plane DC field 
($H_{\rm dc}>4700$\,Oe), the magnetization becomes 
fully polarized along the $z$-direction. We 
took $H=1.02H_c$ in the present simulation.

\begin{figure}[tb]
\begin{center}
   \includegraphics[width=72mm]{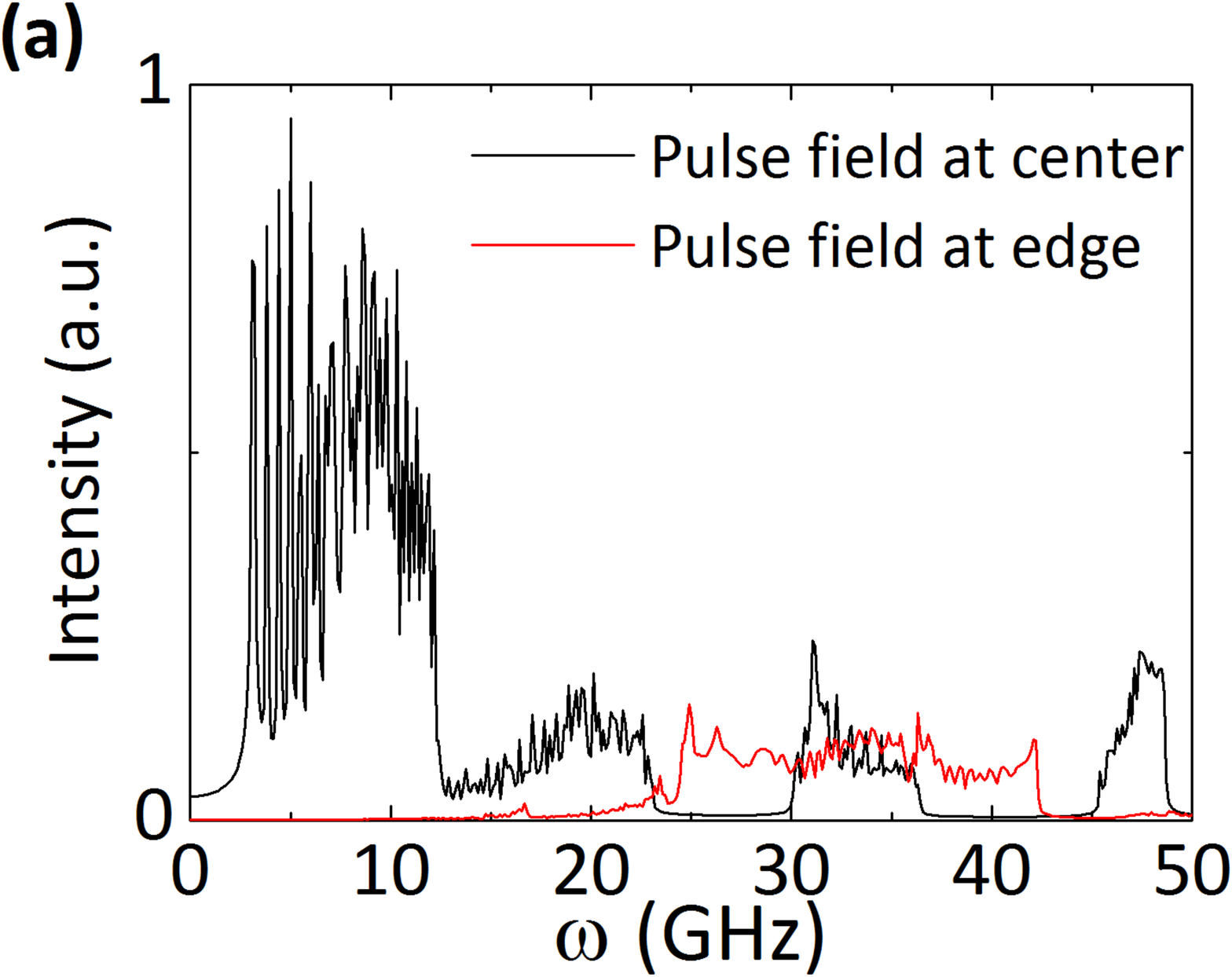}

   \includegraphics[width=40mm]{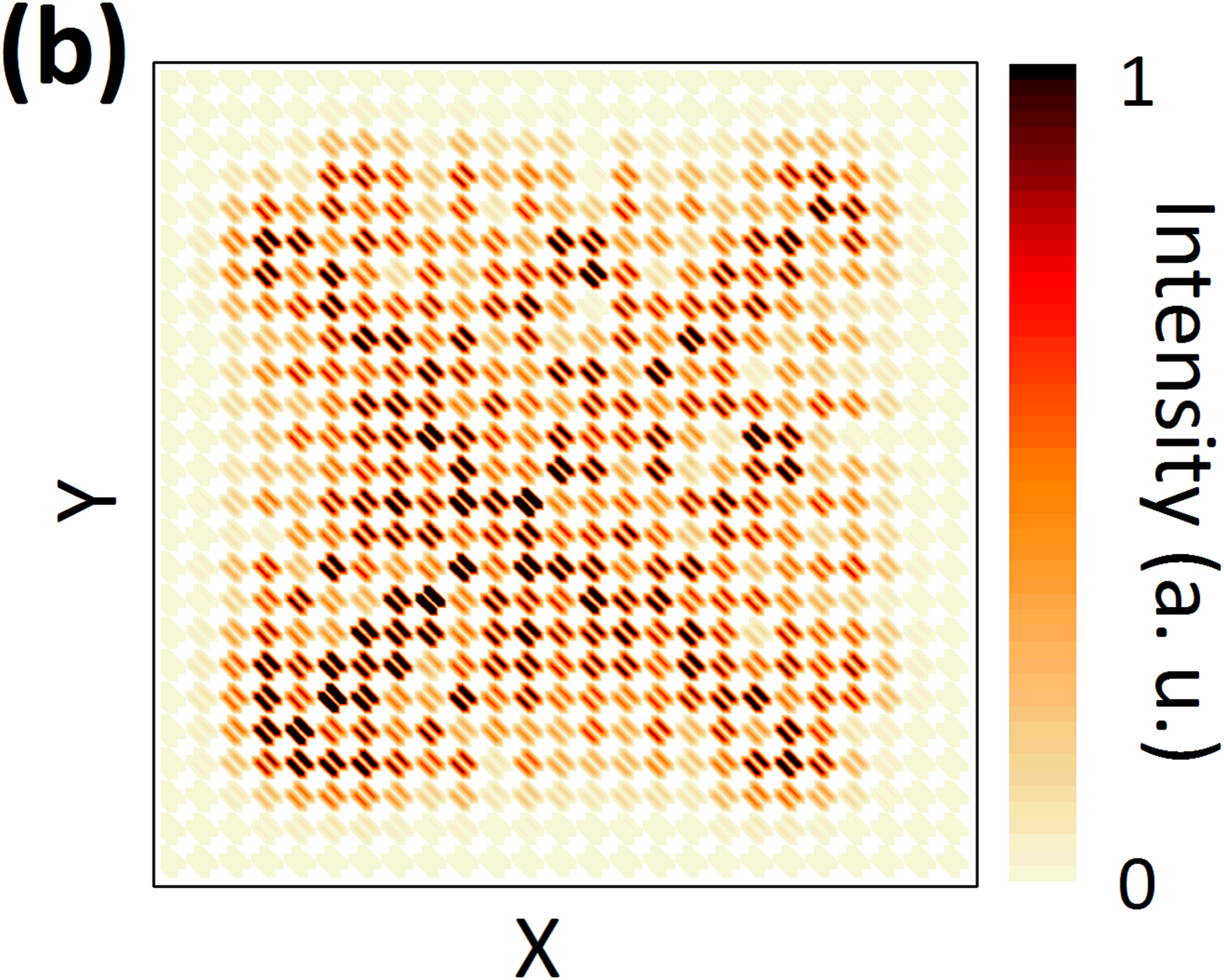}
   \includegraphics[width=40mm]{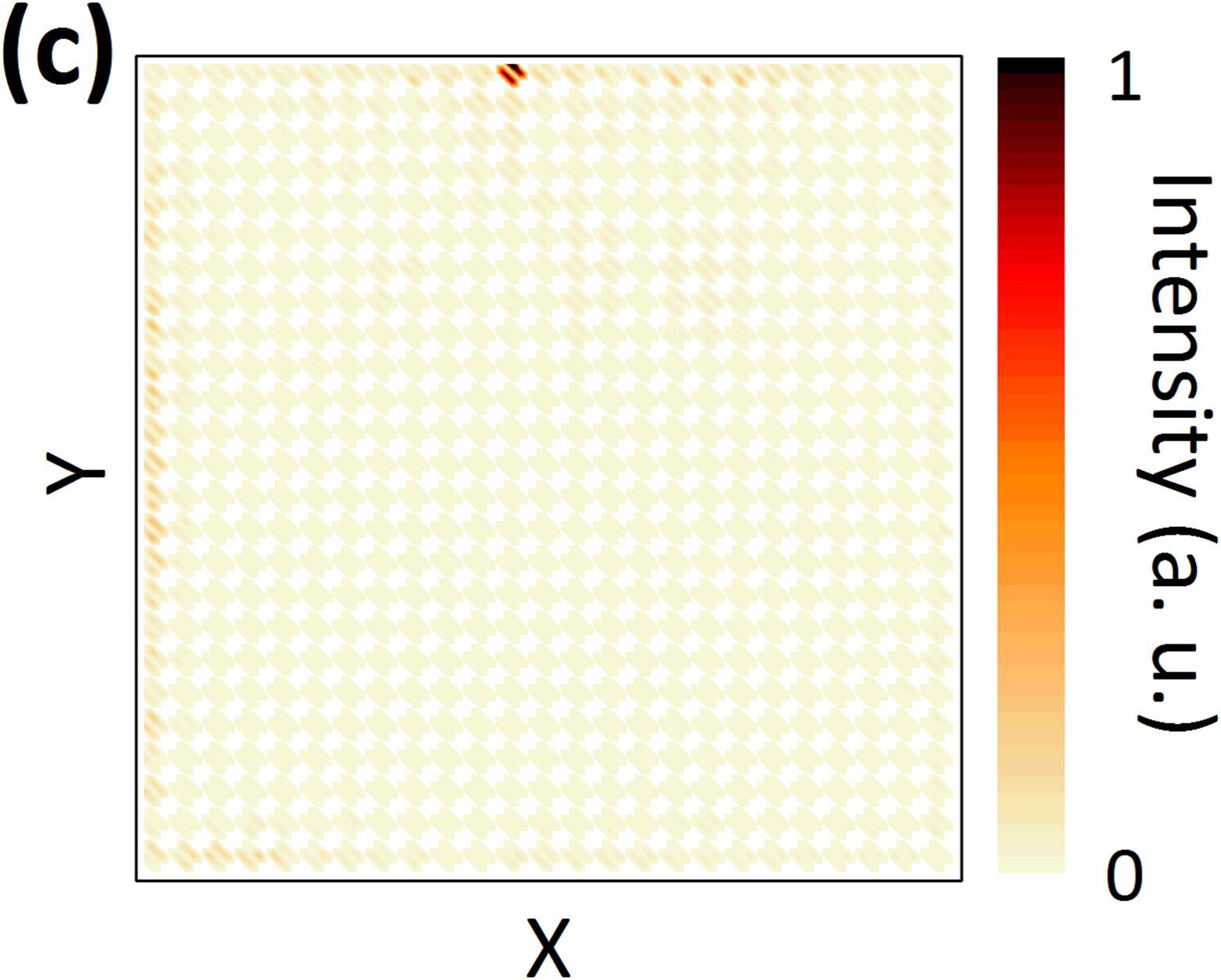}

   \includegraphics[width=40mm]{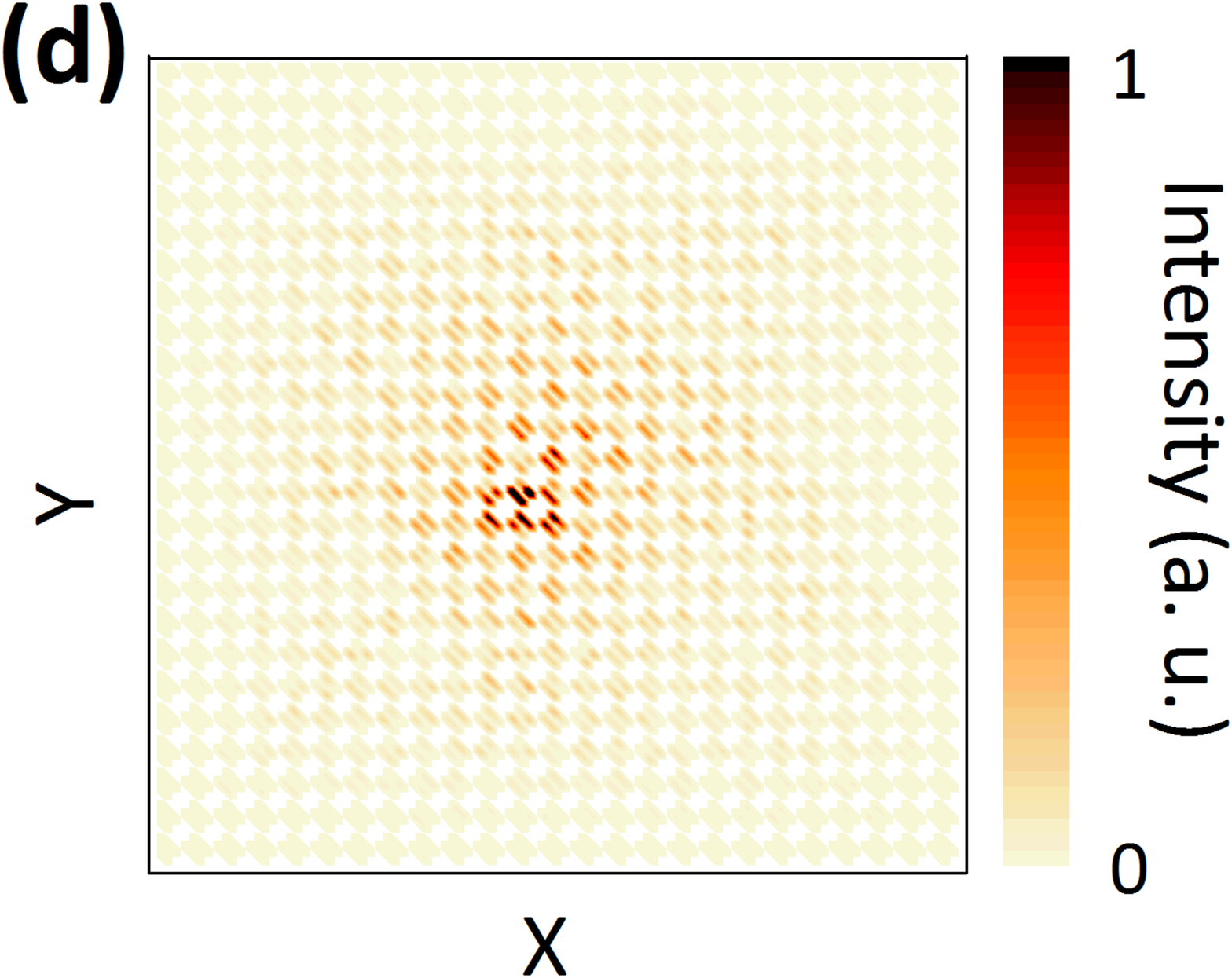}
   \includegraphics[width=40mm]{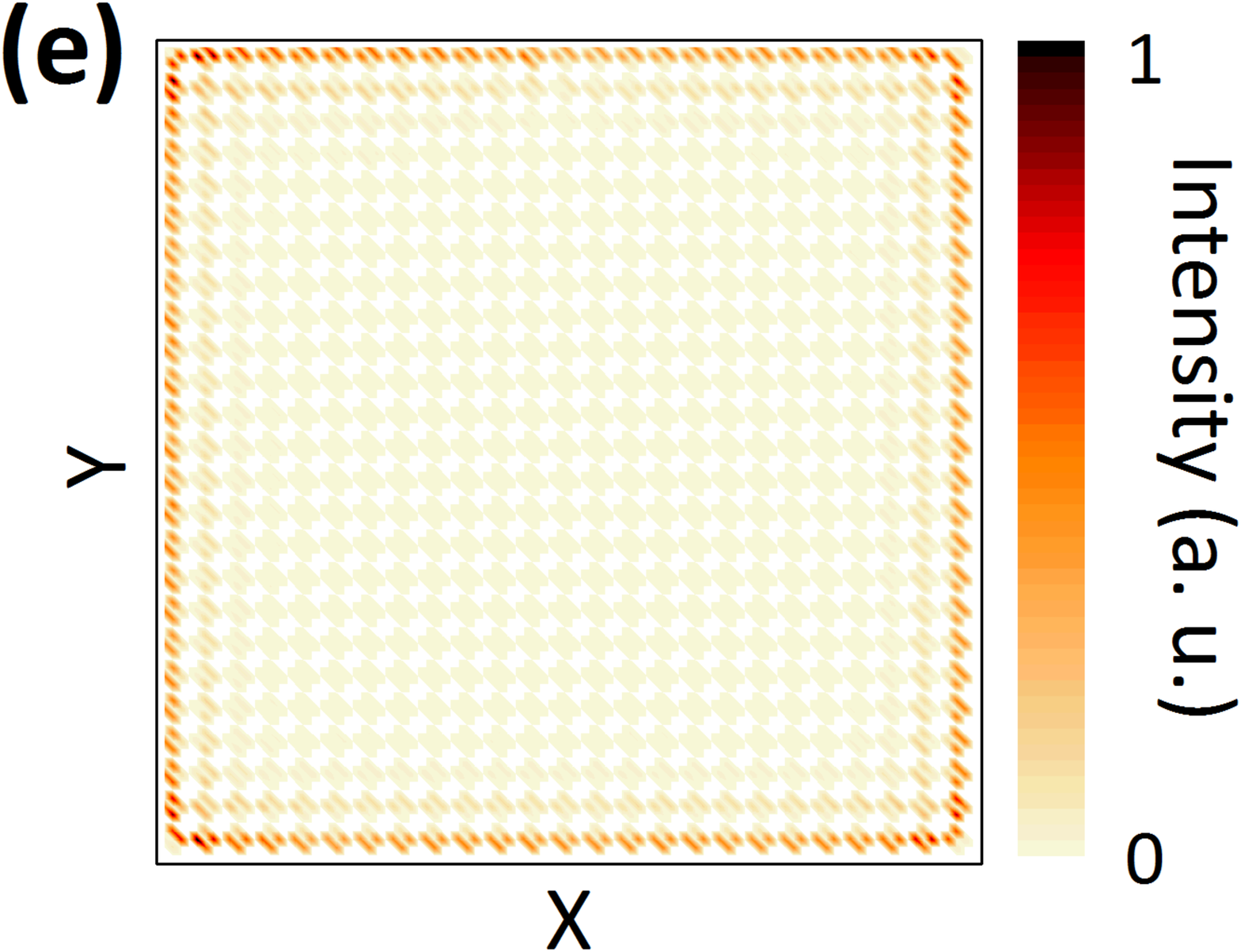}

   \includegraphics[width=40mm]{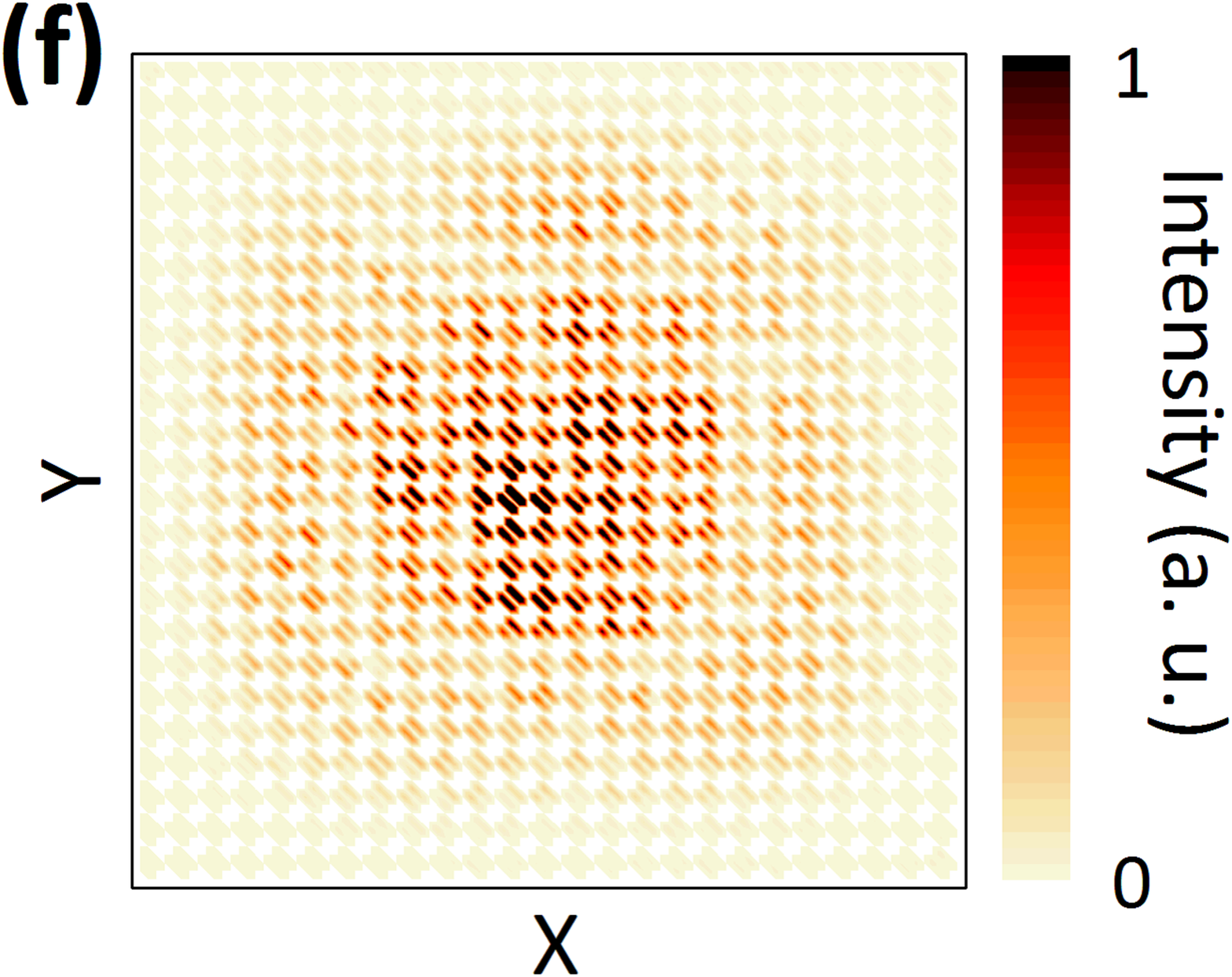}
   \includegraphics[width=40mm]{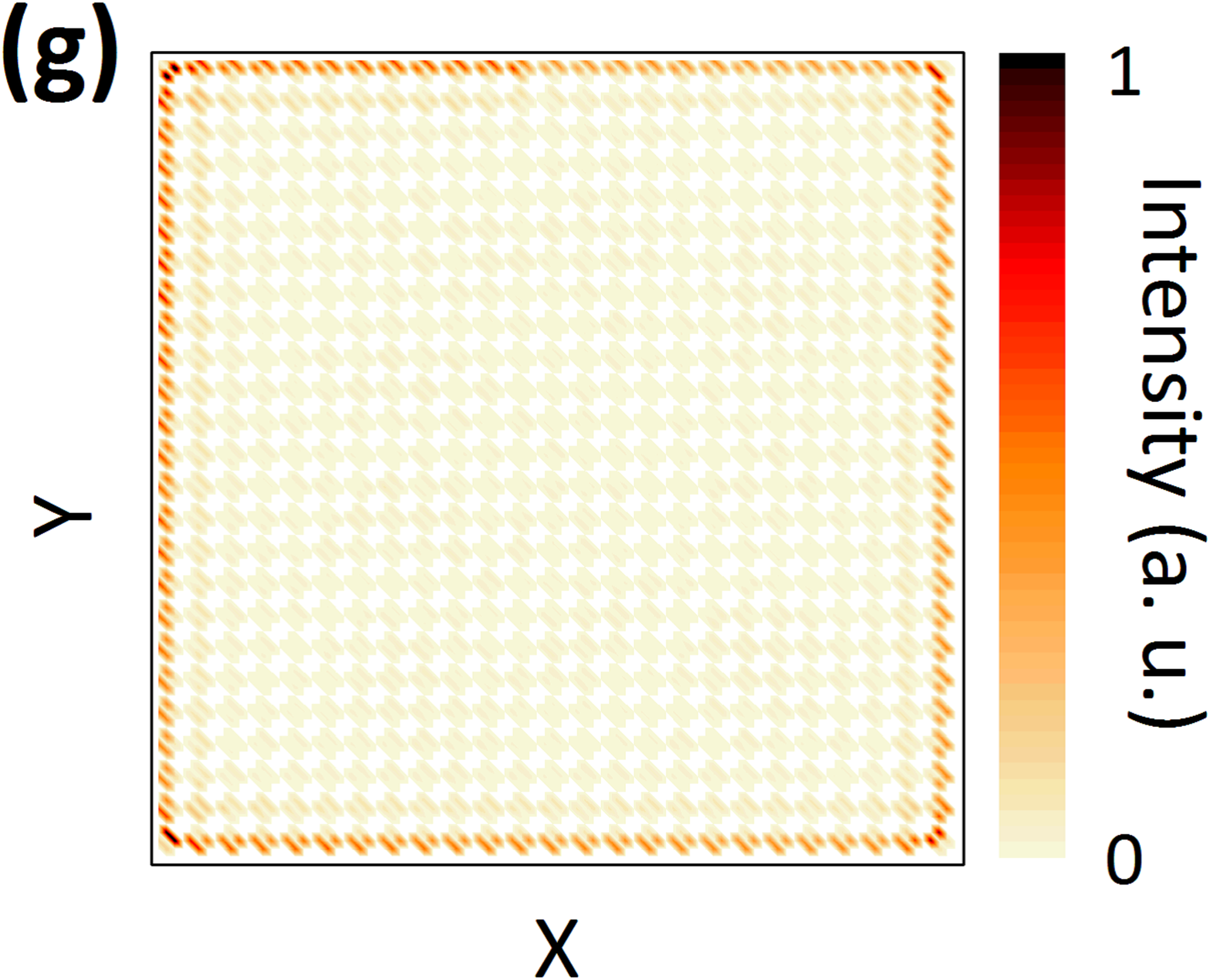}

\caption{ (Color online) Fourier power spectra of magnetization 
dynamics. (a) Frequency dependence of the intensity 
of spin-wave excitations (see text). (b)-(g) Spatial 
distribution of the intensity, which is obtained 
by the application of the pulse field at center (b,d,f) and 
the edge (c,e,g); (b,c) $\omega=10$GHz, (d,e) 
$\omega=29$ GHz, and (f,g) 
$\omega=31$ GHz.}
\label{fig:I_w}
\end{center}
\end{figure}

In order to excite spin wave modes in a broad frequency 
range, we apply a pulsed magnetic field within a plane 
with its pulse time $t_{\rm p}=1$ ps and 
its amplitude $H_{\rm p}=1.0 \times 10^{-4}$ Oe.
The pulse is applied locally 
at the center and around an edge of the system 
for the purpose of exciting volume modes 
and edge modes, respectively. After calculating a 
time evolution of the magnetization in the system,  
we take a Fourier transformation of the transverse moment, 
$m_{+}(X,Y,t)\equiv m_x(X,Y,t)+
{\rm i}m_y(X,Y,t)$, with respect to time; 
\begin{align}
s_{+}(X,Y,\omega) \equiv \sum_{j=0}^{n-1} m_{+}(X,Y,j \Delta T) 
\exp\left( 2\pi {\rm i} \omega j \Delta T \right) \label{5-1}
\end{align}
with $\Delta T=50$ ps and $n=1024$.    
The frequency power 
spectrum, $\sum_{X,Y}  \big|s_{+}(X,Y,\omega)\big|$,  
obtained by the pulse at the center 
and that by the pulse at the edge are shown 
in Fig.~\ref{fig:I_w}(a) separately.
Spatial distributions of spin  
wave excitations, $\big|s_{+}(X,Y,\omega)\big|$, 
for each case with different frequencies $\omega$ 
are shown  in Figs.~\ref{fig:I_w}(b)-(g). From them, one can see 
that the spin-wave volume modes and edge modes are 
selectively excited, depending on whether the initial 
pulse field is applied at the center 
or at the edge respectively. In the case of the pulse 
field at the center, we observe two band gaps 
for volume modes; one from  
$24$GHz to $30$GHz and the other 
from $37$GHz to $46$GHz. In the case of 
the pulse at the edge, we observed spin-wave 
edge modes mainly from $24$GHz to $42$GHz.

\begin{figure}[tb]
\begin{center}
   \includegraphics[width=80mm]{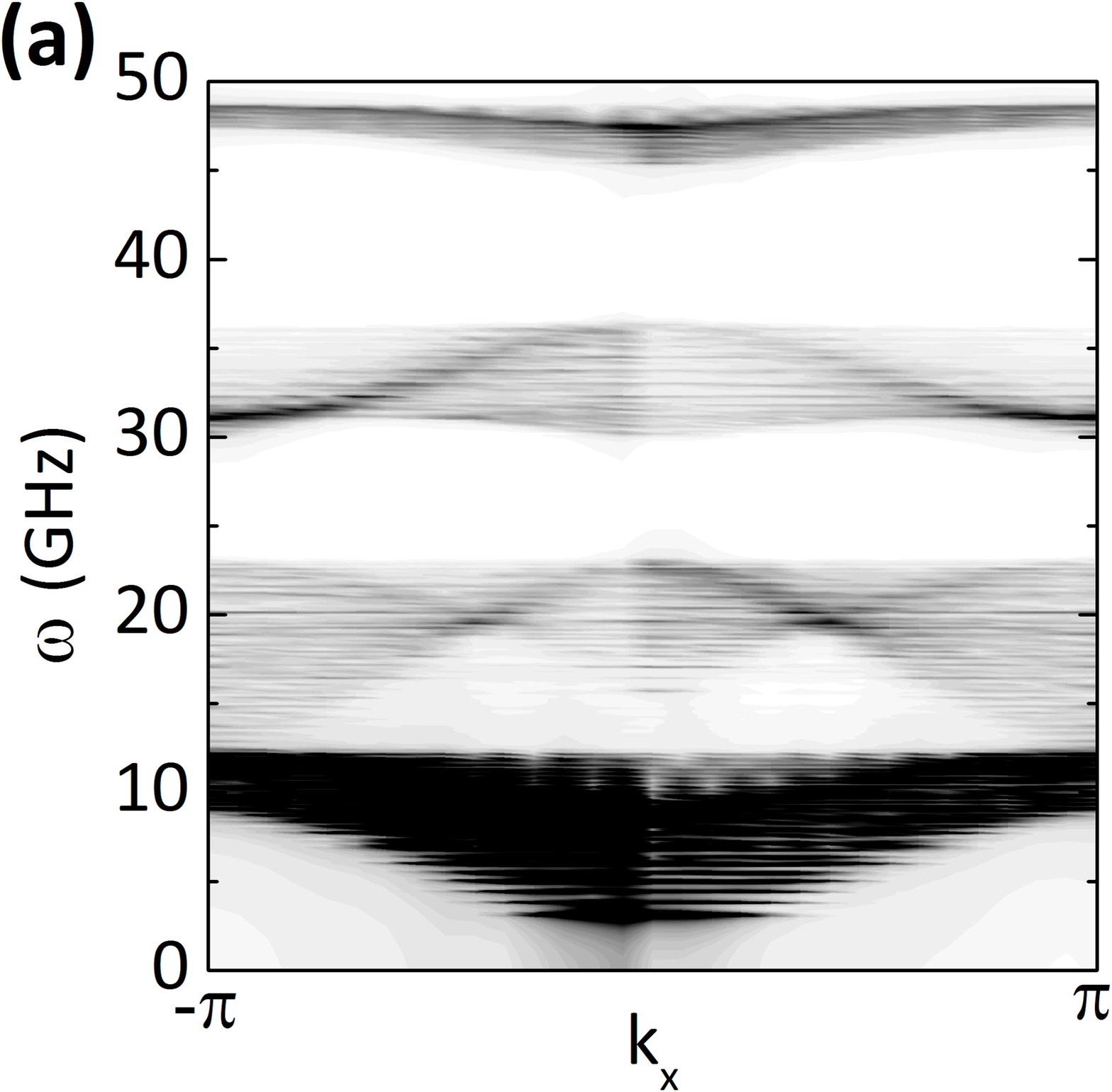}

   \includegraphics[width=42mm]{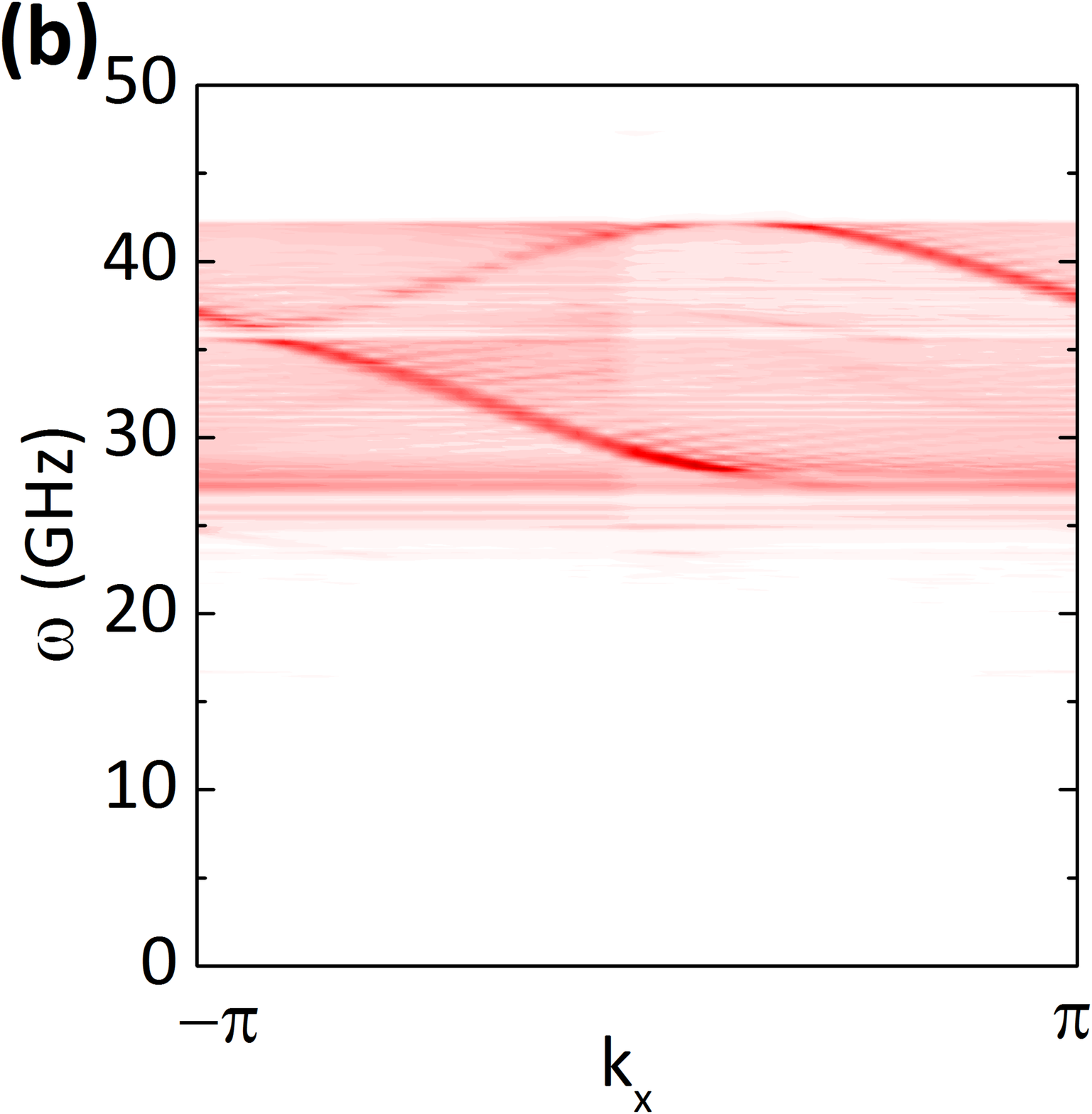}
   \includegraphics[width=42mm]{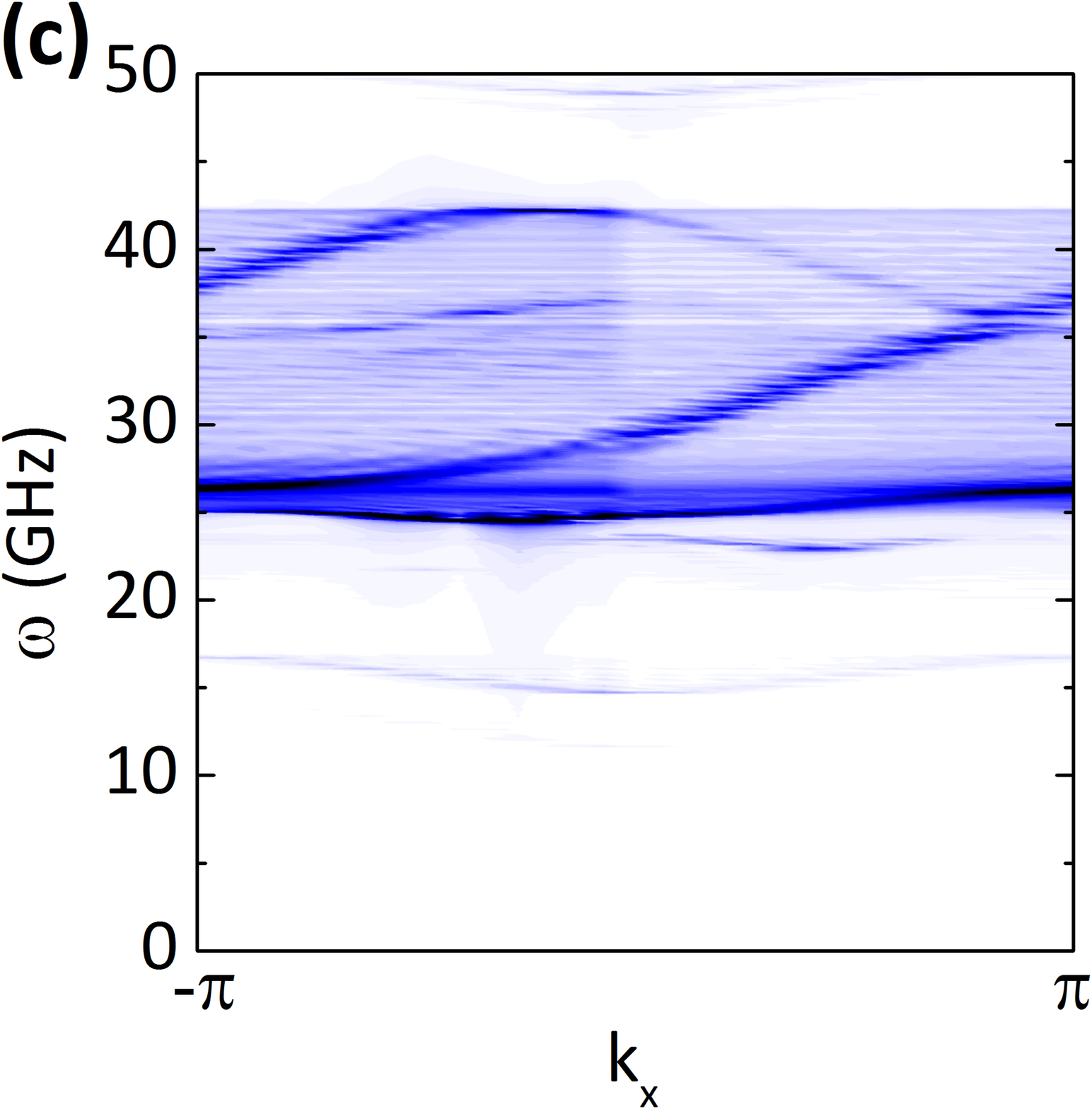}
\caption{ (Color online) Dispersion relations of spin wave modes. (a) 
Dispersion relation obtained by the application of the pulse field 
at the center. (b),(c) 
Dispersion relation obtained by the pulse at the edge. The Fourier 
transformation is taken only over the upper side ($Y>L/2$) for (b)
and over the 
lower side ($Y<L/2$) for (c).}.
\label{fig:dispersion}
\end{center}
\end{figure}

A key feature of proposed chiral spin-wave edge mode is a 
unidirectional propagation of spin wave densities, 
which is clarified by its frequency-wavelength dispersion relation. 
To obtain such a dispersion relation, we next take a Fourier 
transformation of the transverse moment with 
respect to both space and time. 
In order to compare the result 
with Figs.~\ref{fig:b},\ref{fig:c}, we integrate 
the amplitude of the Fourier component  
with respect to the $Y$-component 
of the momentum;
\begin{align}
A(k_x,\omega) &= \sum_{k_y} \Big| s_{+}(k_x,k_y,\omega)\Big|, \! \nn \\
s_{+}(k_x,k_y,\omega) &= \sum_{X,Y} s_{+}(X,Y,\omega) \!\ 
\exp \left({\rm i}k_x X\right)\exp \left({\rm i} k_y Y\right) \!. \nn 
\end{align} 
A contour plot of $A(k_x,\omega)$ as a function of 
$k_x$ and $\omega$ gives a dispersion relation for 
spin-wave modes. Figure~\ref{fig:dispersion}(a) 
shows the dispersion relation for 
the case of applying the pulse field at the center. 
It resembles those for spin-wave volume 
modes obtained in the preceding model calculation 
at the same parameter regime  
(Fig.~\ref{fig:a}(e)). Figures~\ref{fig:dispersion}(b) and (c) 
show the dispersion relations for the case of applying the 
pulse field at the edge. To clarify propagation directions 
of those two spin-wave edge modes running along the 
opposite boundaries of the system, we take the Fourier 
transformation only over the upper (or lower) side of 
the sample $L/2<Y<L$ (or $0<Y<L/2$); the one for the upper 
side is shown in Figure~\ref{fig:dispersion}(b), while the one 
for the lower is in Figure~\ref{fig:dispersion}(c). 
Both figures clearly indicate the existence of 
two counter-propagating chiral dispersions, 
each of which runs across any line of  
$\omega=\omega_0$ ($25{\rm GHz}<\omega_0<35{\rm GHz}$) 
once and only once. The results also 
suggest an existence of another spin-wave edge mode 
from $35$GHz to $42$GHz, which has a quasi-parabolic 
dispersion. Both of these edge modes in combination 
with volume modes shown in Fig.~\ref{fig:dispersion}(a) 
are consistent with the dispersion relations for spin-wave 
modes obtained in the preceding model calculation 
at the same parameter regime (Fig.~\ref{fig:c}(A),(A-1)).

\begin{figure}[tb]
\begin{center}
   \includegraphics[width=42mm]{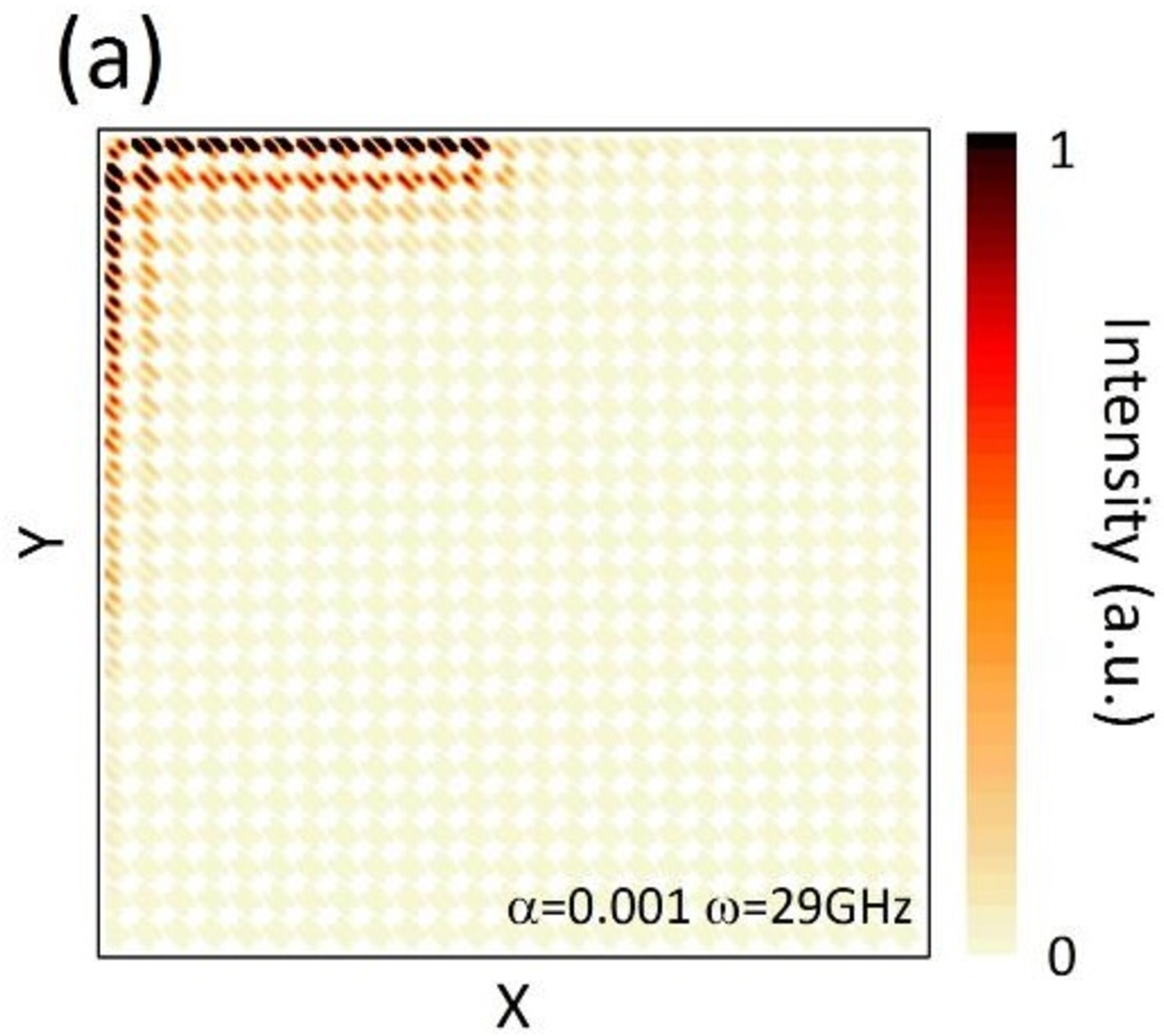}
   \includegraphics[width=42mm]{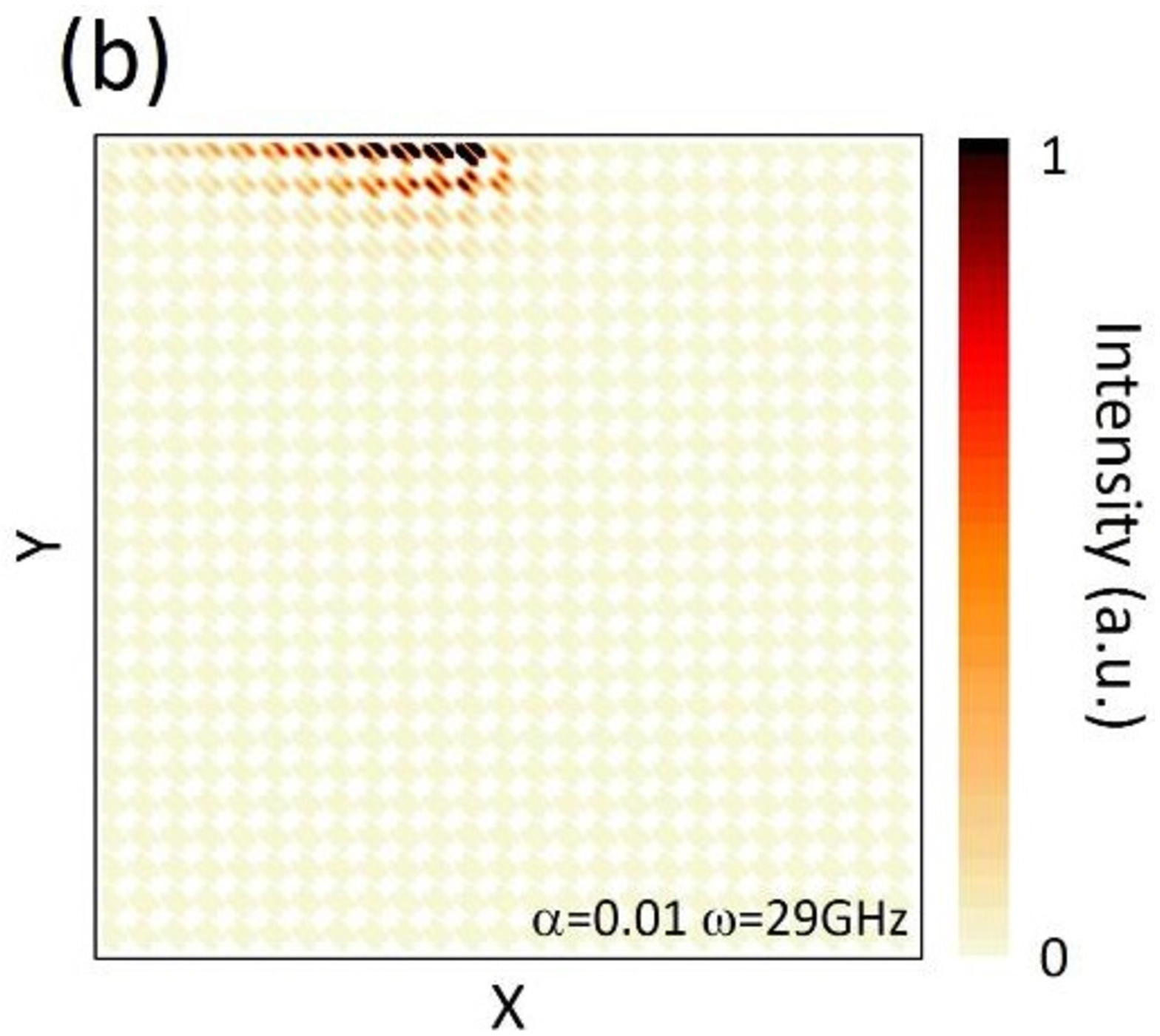}
\caption{(Color online) Spatial-resolved Fourier power 
spectra of magnetization dynamics in the 
presence of stronger dissipation. Spatial 
distribution of the intensity at $\omega=29$GHz, 
which is obtained by the application of the pulse 
field at the edge. (a) $\alpha=0.001$ (b) $\alpha=0.01$.}
\label{fig:dissipation}
\end{center}
\end{figure}

When the Gilbert damping coefficient becomes 
larger, unidirectional propagations of spin density 
along the chiral spin-wave edge mode decay faster. 
Fig.~\ref{fig:dissipation} shows spatial distributions  
of Fourier power spectra of magnetization dynamics, 
$|s_{+}(X,Y,\omega)|$, in the presence of larger 
Gilbert damping term ($\alpha=0.001,0.01$), where 
the initial pulse field is applied at the edge 
(Fig.~\ref{fig:system}) and the frequency 
is chosen within the band gap ($\omega=29$GHz). 
The results suggest that the coherence length 
is roughly $25$ unit cell size ($500$nm) for $\alpha=0.001$ 
and $8$ unit cell size ($160$nm) for $\alpha=0.01$.

\section{summary, disucssions and open issues}
In this paper, we introduced two simple 
magnetic thin-film models, in which 
ferromagnetic nanoislands on periodic arrays  
are coupled with each other via magnetic dipolar 
interaction. Under the field applied 
perpendicular to the two-dimensional plane, 
spin-wave excitations in the systems 
have a chiral spin-wave edge mode localized at 
the boundaries of the systems, whose 
dispersion runs across a band gap for 
spin-wave volume modes. The sense 
of the rotation of the chiral edge mode 
is determined by a sign of the Chern 
integer for a spin-wave volume-mode band 
below the band gap.

To have volume-mode bands 
with finite Chern integer, we generally 
need multiple-band degree of freedom within 
a unit cell. To this end, we considered two periodic 
arrays of ferromagnetic particles; decorated 
square-lattice model and honeycomb-lattice 
model.  For the decorated square-lattice 
model, we observed that, on increasing the out-of-plane 
field, there appears a sequence of band 
touchings between pairs of neighboring volume-mode  
bands. Owing to these band touchings, 
the Chern integers for the volume modes change their 
signs and, concomitantly, the chiral edge   
mode changes its sense of rotation from clockwise 
to anticlockwise or vice versa. For the decorated 
honeycomb-lattice model, we observed a finite band 
gap between the lowest spin-wave volume-mode band 
and second lowest spin-wave band which are 
connected by a chiral dispersion of an edge  
mode. Though its sense of rotation being unchanged 
by the strength of the field in the honeycomb lattice case, 
the gap and the chiral edge 
mode persists for a sufficiently large field. 

To interpret these results, 
we next construct tight-binding descriptions 
for the linearized Landau-Lifshitz equation, 
in which atomic orbitals such as $s$-wave, 
$p_{\pm}$-wave and $d_{x^2-y^2}$-wave orbitals 
are introduced within 
each unit cell. Among other, complex-valued characters 
in the $p_{\pm}$-wave orbitals break both the 
time-reversal symmetry and mirror symmetries 
of the models. These symmetry 
breakings lead to a non-zero Chern integer for  
spin-wave volume-mode 
bands and associated chiral spin-wave edge 
modes. Using this tight-binding model, we argue 
that the level inversions among different atomic 
orbital levels give rise to the so-called {\it inverted} spin-wave 
bands with non-zero Chern integers.  Our tight-binding 
analysis for the square-lattice model 
gives quantitative criteria for the emergence of 
finite field ranges within which spin-wave 
volume-mode bands have non-zero Chern integers. 

For the decorated honeycomb lattice model, we employ 
a perturbation analysis, starting from the large field limit. 
The analysis suggests that the effective Hamiltonian 
in the large field 
limit always respects time-reversal symmetry 
and the hexagonal symmetry within the order of 
${\cal O}(1)$. Due to the mirror operations 
in the hexagonal symmetry, the lowest two 
spin-wave bands form gapless Dirac cone spectra at 
two inequivalent $K$-points. Once ${\cal O}(1/H)$-order 
corrections are included, however, the time-reversal 
symmetry is lost and hexagonal symmetry reduces 
to its abelian subgroup having no mirror symmetries. As a 
result, the gapless Dirac-cone spectra acquire 
a finite mass of the order of ${\cal O}(1/H)$, 
which leads to non-zero Chern integer 
for the two lowest spin-wave bands.  This argument 
explains why the spin-wave volume-mode bands 
with non-zero Chern integers and associated chiral  
spin-wave edge mode persists in a very wide range of the 
field in the decorated honeycomb lattice model. 

Since a state-of-the-art sample production does not necessarily 
guarantee perfect periodic structurings, considering 
disorder effects associated with the lattice periodicity are 
experimentally relevant, which can be speculated 
from well-established knowledges of integer Quantum 
Hall physics.~\cite{Halperin,Aoki-Ando,QHE,onoda-nagaosa2} 
The effects are two-folded. When the strength of 
the disorders is smaller than a characteristic frequency scale 
of the band gap, those volume modes near the band gap 
become localized due to the disorders, while chiral edge 
mode itself is free from these weak disorders. As a result, the 
frequency regime for the chiral spin-wave edge mode  
becomes even wider than that in the clean limit. 
When the strength of the disorders is increased to be 
larger than the scale of the band gap in the clean limit, 
however, the `mobility gap' closes and reopens. 
After the reopening the gap, the topological chiral 
edge mode disappears.~\cite{Aoki-Ando,QHE,
onoda-nagaosa2} 
The proposed chiral spin-wave edge mode is also robust 
against the boundary shape; the edge modes persist in 
almost arbitrary shapes of the boundary, provided that 
the edge mode in the boundary has no interference 
with the other mode running along the opposite 
sample boundary.~\cite{Halperin,Hatsugai} 

It is also a non-trivial issue whether 
submicrometer-scale ferromagnetic islands 
behave as a single spin or not. In preceding 
experimental systems mentioned before,~\cite{asi,CARoss}  
non-isotropic shapes of ferromagnetic islands give  
rise to strong magnetic dipolar anisotropies, 
forcing all the spins in each island 
to point along a same direction. In our 
model calculations, magnetic anisotropies within each 
island are not included from the outset. It is  
interesting to include these magnetic dipolar 
anisotropies into the present 
Landau-Lifshitz equation phenomenologically 
as the `single-ion' type magnetic anisotropies. 
It is also equally likely that a ferromagnetic island has 
a couple of low-frequency relevant modes having 
different spin textures within the island. 
Such modes can be also utilized as a kind of 
`atomic orbital', so that a system with only one ferromagnetic 
island within a unit cell could also have a chance 
to provide a volume-mode bands with finite Chern integers. 
Exploring such systems is, however, beyond the scope 
of the present paper and we leave them for future 
open issues.~\cite{later} 

\begin{acknowledgments}
We would like to thank Y. Suzuki, 
S. Miwa and J. Feng for helpful discussions. 
This work is supported in part by Grant-in-Aids  
from the Ministry of Education,
Culture, Sports, Science and Technology of Japan
 (No.~21000004, 22540327, 23740225 and 
24740284) and by Grant for Basic Scientific 
Research Projects from the Sumitomo Foundation.  
\end{acknowledgments}  

\end{document}